\documentclass{aa}
\usepackage[varg]{txfonts}

\usepackage{natbib}
\bibpunct{(}{)}{;}{a}{}{,} 
\usepackage[T1]{fontenc}
\usepackage{ae,aecompl}
\usepackage{graphicx}
\usepackage{xspace}
\usepackage{amsmath}
\usepackage{amssymb}
\usepackage{bbold}
\usepackage{epsfig}
\usepackage{pdflscape}
\usepackage{caption}
\usepackage{subcaption}
\usepackage{verbatim}
\usepackage{amsfonts}
\usepackage{hyperref}
\usepackage{color}
\usepackage{float}
\usepackage{url}
\usepackage{booktabs}
\usepackage{fancyhdr}
\usepackage{lipsum}
\usepackage{multirow}
\usepackage{grffile}
\usepackage{multirow}
\usepackage{algorithm}
\usepackage{algpseudocode}
\usepackage{siunitx}

\usepackage{geometry}
\newcommand{\ie}{i.\,e.}

\newcommand{\eg}{e.\,g.}
 
\newcommand{\mcala}{${A}$}
\newcommand{\pl}{\textsc{Planck}\xspace}
\newcommand{\pb}{\textsc{Polarbear}\xspace}

\newcommand{\cmb}{CMB\xspace}
\newcommand{\mbd}{M_{BD} }
\newcommand{\xpure}{\texttt{X$^{2}$PURE}}
\newcommand{\md}{M_{2l} }

\newcommand\gpedit[1]{ }

\begin{document}
\title{Iterative map-making with two-level preconditioning for polarized Cosmic Microwave Background data sets}
\subtitle{A worked example for ground-based experiments}
\titlerunning{Iterative map-making with two-level preconditioning for polarized CMB data sets}
\authorrunning{G. Puglisi et al.}
\author{ Giuseppe Puglisi\inst{1} \thanks{\email{giuspugl@sissa.it}}  \and Davide Poletti\inst{1}   \and Giulio Fabbian  \inst{2,1} \and Carlo Baccigalupi \inst{1,3} \and{Luca Heltai}\inst{1}\and  Radek Stompor\inst{4}  }

\institute{
SISSA- International School for Advanced Studies, Via Bonomea 265, 34136 Trieste, Italy
  \and  
  Institut d'Astrophysique Spatiale, CNRS (UMR 8617), Univ. Paris-Sud, Universit\'{e} Paris-Saclay, b\^{a}t. 121, 91405 Orsay, France
\and 
  INFN-National Institute for Nuclear Physics, Via Valerio 2, I-34127 Trieste, Italy
  \and AstroParticule et Cosmologie, Univ Paris Diderot, CNRS/IN2P3,CEA/Irfu, Obs de Paris, Sorbonne Paris Cit\'{e}, France
  } 

\date{\today }

\abstract {An estimation of the sky signal from  streams of Time Ordered Data (TOD) acquired by the Cosmic Microwave Background (\cmb) experiments is one of the most important steps in the context of \cmb data analysis referred to as the map-making problem. The continuously growing \cmb data sets render the \cmb map-making problem progressively more challenging in terms of computational cost and memory in particular in the context of ground based experiments with their operational limitations as well as the presence of contaminants. }
{We study a recently proposed, novel class of the Preconditioned Conjugate Gradient (PCG) solvers
which invoke two-level preconditioners in the context of the ground-based CMB experiments. We compare them against the PCG solvers commonly used in the map-making context considering their precision and time-to-solution.} 
{We compare these new methods on realistic, simulated data sets reflecting the characteristics of current and forthcoming \cmb ground-based experiment.
We develop an embarrassingly parallel, \emph{divide-and-conquer} implementation of the approach where each processor performs a sequential map-making for a subset of the TOD. }%
{We find that considering the map level residuals the new class of solvers permits achieving  tolerance of up to 3 orders of magnitude better than the standard approach, where the residual level often saturates before convergence is reached. This often corresponds to an important improvement in the precision of the recovered power spectra in particular on the largest angular scales. The new method also typically requires fewer iterations to reach a required precision and thus shorter runtimes for a single map-making solution. However, the construction of an appropriate two-level preconditioner can be as costly as a single standard map-making run. Nevertheless, if the same problem needs to be solved multiple times, e.g., as in Monte Carlo simulations, this cost has to be incurred only once, and the method should be competitive not only as far as its precision but also its performance is concerned.}
{}
\maketitle
\section{Introduction}
Over the last decades, several experiments have looked into the CMB polarization anisotropies aiming at  discovering a stochastic background of gravitational waves produced during the inflationary phase of our Universe encoded in the $B$-modes, \ie\ the divergence-free pattern in \cmb polarization. Indeed, the amplitude of the CMB $B$-mode polarization anisotropies at the scales larger than $1$ degree, conventionally parameterized with a \emph{tensor-to-scalar ratio}, $r$, is thought to be directly related to the energy scale of inflation ($\sim 10^{16}$ GeV). These \emph{primordial} $B$-modes have not been detected yet and further progress in both the control of the diffuse polarized emission from our Galaxy (involving widely the microwave frequency regime \citep{PhysRevLett.114.101301}) and in the sensitivity of the experimental set-ups is necessary in order to reach such a goal. At the sub-degree angular scales,  $B$-modes are produced by the gravitational lensing due to  large scale structures intervening along the photon path travelling towards us.  Evidence for these lensing $B$-modes was first provided via cross-correlation of the CMB polarization maps with the cosmic infrared data \citep{2013PhRvL.111n1301H, 2014PhRvL.112m1302A} and via constraining the small-scale $B$-mode power \citep{2014ApJ...794..171T} and they have been since then characterised with increasing accuracy \citep{2016arXiv161002360L,sptpol2015,Bicep2Keck2016, ThePOLARBEARCollaboration2017}.
While these past experiments have observed the microwave sky with arrays of thousands of detectors often focusing on small sky patches, the forthcoming \cmb experiments are planned to observe bigger patches with at least tens of thousands of detectors, producing as a result, Time Ordered Data (TOD) including tens and hundreds of billions of samples.  

The recovery of the sky signal from these huge, noisy time streams, a process called \emph{map-making}, represents one of the most important steps in \cmb data analysis and, if the detector noise properties and scanning strategy are known,  map-making becomes a linear inverse problem.
The Generalized Least-Squares (GLS) equation provides an unbiased solution to map-making for an arbitrary choice of weights given by a symmetric and positive definite matrix~\citep{1997PhRvD..56.4514Tb}.
Moreover, if we consider as the weights the inverse covariance of the time-domain noise, the GLS  estimate is also a minimum variance and a maximum likelihood solution to the problem. However, a computation of the solution in such a case may require either an explicit factorisation of a huge, dense matrix~\citep{1997PhRvD..56.4514Tb,1999astro.ph.11389B,2002PhRvD..65b2003S}  or an application of some iterative procedure~\citep{1996astro.ph.12006W,1999ApJ...510..551O, 2001A&A...374..358D,2005A&A...436.1159D,2010ApJS..187..212C}. These latter involve typically several matrix-vector multiplications at each iteration step.
\noindent  What makes the map-making problem particularly challenging are the sizes of the current and forthcoming CMB data sets which are directly related to the number of floating point operations (flops) needed to achieve the solution and to the memory requirements due to the sizes of the arrays required for it. 
 Both these factors set the requirements on computational resources and indeed many current \cmb data analysis pipelines opt for  massively parallel computing platforms. However, even in such circumstances efficient algorithms are necessary to ensure that the analysis can be indeed performed. The computational complexity of the algorithms involving  an explicit matrix inversion is $ \mathcal{O}(N_p ^3)$ flops,  where $N_p$ is the  number of pixels in the map, and therefore they are suitable only for the cases when the estimated sky maps do not involve many sky pixels. 
However, whenever feasible the direct approaches can yield high-precision, unbiased estimates of the sky signal  \citep[e.g,][for a recent example]{Poletti2016}. However, the next generations of the ground experiments, \cmb-Stage III ~\citep{arnold2014, advactpol, spt3g} and IV~\citep{2016arXiv161002743A}), are expected to observe significant fractions of the entire sky with high resolution and thus resulting in maps with  $N_p \simeq {\cal O}(10^6)$,  rendering the direct approaches prohibitive even for the largest forthcoming supercomputers.

In this context iterative methods have offered an interesting alternative.
%
%
They involve algorithms within the class of  \emph{Krylov} methods \citep[e.g.,][and references therein]{Golub:1996:MC:248979}, which avoid the explicit inversion of the linear system matrix by constructing an approximate solution which is iteratively improved on. The  computational complexity of such methods is mostly driven by  matrix-vector products, which need to be performed repeatedly on each iteration. These require at most ${\cal O}(N_p^2)$ flops and can be performed at much lower cost in the specific case of the CMB map-making (see Sect.~\ref{sec:pcg}), where such matrix-vector products can be computed matrix-free, i.e., without ever assembling explicitly the system matrix in  memory~\citep{2010ApJS..187..212C}. To date, most of \cmb iterative solvers have been based on the  \emph{Conjugate Gradient} (CG) method applied to a preconditioned system of map-making equation and involved a simple block-diagonal preconditioner (see Eq.~\eqref{eq:mbd}). While these solvers performed usually very well~\citep[e.g., ][and references therein]{2009A&A...493..753A, 2010ApJS..187..212C}, the anticipated data sets motivate a search for better, more efficient algorithms~\citep{Grigori2012, 2014JCAP...10..007N,2014A&A...572A..39S, huffenberger2017}.

In this paper we apply the methodology proposed by \citet{2014A&A...572A..39S} to a reconstruction of maps from simulated data of a modern, ground-based \cmb experiment. This new class of approaches involves constructing and applying a more involved, two-level preconditioner. Our simulations are informed by the experiences derived from the deployment and analysis of the \pb experiment whose results from the first   two seasons of data have been recently published in \citet{ThePOLARBEARCollaboration2017}. In Sect.~\ref{sec:mm}, we briefly introduce the formalism of the  map-making problem in presence of time domain filtering operators typical for a ground based \cmb experiment. In Sect.~\ref{sec:pcg}, we describe the iterative approach and the two different methodologies adopted in the analysis. In Sect.~\ref{sec:degen} we further describe how the filters introduce degeneracies on the estimation of the maps. \gpedit{The main results of this paper are presented in Sect.~\ref{sec:results}, where we show a close comparison of the performances of the commonly-used block-diagonal and the two-level preconditioners applied onto realistic simulated data sets. These include signal as well as sky coverage and noise properties representative of the forthcoming generation of \cmb polarization experiments. Finally, in Sect.~\ref{sec:conclusions} we summarize our main results and conclusions.}

\section{Map-making in \cmb ground-based experiments} \label{sec:mm}

The input data of the map-making procedure are the calibrated TODs collected in a single time-domain vector $d$ of size $N_t$ containing all measurements performed during a certain period of time by all the detectors of a CMB experiment. The measurements can be modelled as the sum of an astrophysical signal and  measurement noise, ${n}_t$. The astrophysical contribution to a measurement taken at time $t$ is given by the sky signal, $s_p$, in pixel $p$ observed at time $t$ and which is already convolved with the instrument response, assumed hereafter to be axially symmetric. The correspondence between the sky pixel, $p$, and the time, $t$, can be encoded by  a sparse and tall ($N_t \times N_p$) matrix, ${P}_{tp}$.  The data model can be then written as: 
\begin{equation}
 {d}_t= {P}_{tp}  {s}_p + {n}_t,
\label{eq:probl}
\end{equation} 
or in the matrix form as,
\begin{equation}
d= Ps + n. \label{eq:datamod}
\end{equation}
Here, $s$ stands for the map to be estimated.

The structure of the pointing matrix encodes the scanning strategy of the \cmb experiment and depends on whether the detectors are sensitive or not to the polarization. In the former case, the sky signal is described by three \emph{Stokes parameters} $I,\,Q,\,U$ in every pixel $p$ of the map, i.e., $s_p=(I_p,Q_p,U_p)$, and a measurement by a polarization-sensitive experiment taken at time $t$ can be written explicitly as, 
\begin{equation}
d_t= I_{p_t} +Q_{p_t}\cos(2 \phi_t) + U_{p_t} \sin (2 \phi_t) + n_t, \label{eq:datamodel}
\end{equation}
where $\phi_t$ is the angle of the detector projected onto the sky coordinates at time $t$. The pointing matrix has in this case three non zero entries per row. 
 We further assume the noise to have vanishing mean $\langle n \rangle =0 $ and defined by the noise covariance matrix $\mathcal{N}$. 

Under these assumptions the map-making is a linear inverse problem of estimating the sky signal, $s$, from the data, $d$, given the data model as in Eq.~\ref{eq:datamod}. This is a linear statistical problem whose solution is provided by a GLS,
\begin{equation}
\hat{s} = (P^\dag   W P)^{-1} P^\dag   W d,  \label{eq:estimator}
\end{equation}
yielding an \emph{unbiased } estimator \citep{1997ApJ...480L..87Ta} for any choice of a positive definite matrix $W$.  In particular, if $W= \mathcal{N}^{-1}$ and the noise is Gaussian distributed the estimator in Eq.~\eqref{eq:estimator} becomes minimum variance. 

\subsection{The filtering operator }
\label{sec:filters}
The raw TODs are often contaminated by some unwanted signals that are not astrophysical in their origin, such as the ground pickup or the atmospheric contributions, or their noise properties display strong, long-term correlations commonly referred as \emph{$1/f$} noise. All these contributions are usually filtered out from the data.

\noindent In such cases the template of the unwanted signal, $T$, is known while its amplitude $y$ is not. What a filtering operation is required to do is to remove a component of the TOD contained by the subspace spanned by the columns of $T$, \ie\,
\begin{eqnarray}
d^{\prime}\ \ &\equiv& (\mathbb{1} - T(T^\dag   T)^{-1} T^\dag  ) d = F_T d \\
F_T T&=&0,\label{eq:filt}
\end{eqnarray}
so that $d' \cdot T_i  = 0$, for any template $T_i$ included as a column of the  \emph{template matrix}, $T$. The most general form of the filtering operator involves also weighting by a full-rank weight matrix, $W$, and reads,
\begin{equation}
d'\equiv (W - WT (T^\dag   WT)^{-1} T^\dag  W) d=F_T d.
\label{eq:filtnoise}
\end{equation}
With the above definition of $F_T$,   it is therefore possible to generalize \eqref{eq:estimator} to ~\citep{Poletti2016} 
\begin{equation}
\hat{s} = (P^\dag  F_T P)^{-1} P^\dag   F_T d. \label{eq:unb}
\end{equation}
Notice that the filtering operator does not change the properties of the estimator in \eqref{eq:unb}. It is still unbiased,  
\begin{displaymath}
\langle \hat{s} -s \rangle = \langle (P^\dag   F_T P)^{-1} P^\dag   F_T n  \rangle =0,
\end{displaymath}
and if we consider $W = \mathcal{N}^{-1}$,  it is minimum variance.

\section{Preconditioned iterative solvers}\label{sec:pcg}

We can rewrite \eqref{eq:unb} as a linear system,
\begin{align}
\left( {P}^\dag    {F_T P}\right) \hat{s}&=  {P}^\dag   { F_T d}, \label{eq:linearsystem} \\ 
&\Downarrow \nonumber \\
Ax&=b,\nonumber 
\end{align}
where ${A}$ is a symmetric and positive definite (SPD) matrix.

\noindent The CG algorithm is particularly attractive for large sparse or structured systems since it references the system matrix \mcala\ only through its multiplication of a vector.  
 The convergence rate of the CG depends on the condition number of the system matrix, $\kappa$, \citep{Golub:1996:MC:248979}, defined as the ratio of the largest to the smallest eigenvalue of a matrix. 
 
\noindent To reduce the condition number of ${A}$,  a \emph{preconditioner} matrix $M_P$ is applied to the linear system so that the condition number of matrix $M_P A $ is smaller. If this is the case, the CG  converges within a smaller number of iterations. This new algorithm is commonly referred as \emph{Preconditioned } CG (PCG) as it solves the preconditioned linear system, 
\begin{equation}
M_P {A} x= M_P b.
\label{eq:pcg}
\end{equation}
\noindent
It can be shown  \citep{Golub:1996:MC:248979} that the PCG convergence rate is strictly related to the condition number of the preconditioned matrix $M_P A$. In fact, after $k$ iterations of the PCG, the magnitude of the error is 
 \begin{equation}
\parallel e^{(k)} \parallel  \equiv\frac{\parallel x -x^{(k)} \parallel}{\parallel x \parallel} \leq \kappa(M_P A) \frac{\parallel r ^{(k)}  \parallel}{\parallel b \parallel},
 \label{eq:errnorm}
 \end{equation}
where $x$ is the true solution to \eqref{eq:linearsystem} and 
\begin{equation}
r ^{(k)}\equiv  b - M_P A x ^{(k)} \label{eq:residuals}
\end{equation}
is the PCG \emph{residual} at the $k$-th step.

\subsection{The Jacobi  Preconditioner}\label{subsec:jacobi}
From \eqref{eq:estimator} we can define the \emph{Jacobi Preconditioner}:
\begin{equation}
\mbd^{-1} \equiv P^\dag \text{diag}(\mathcal{N}^{-1})   P.
\label{eq:mbd}
\end{equation}
This is not only trivial to compute, store and apply to a vector, but it also accounts for some of the eigenstructure of the actual system matrix, which is due to the inhomogeneity of the sky observations. These properties justify why $M_{BD}$ is the most popular and very successful preconditioner used in the current CMB map-making practice. We call it either the \emph{block diagonal} or {\em Jacobi} preconditioner\footnote{We note that a more typical definition of the Jacobi preconditioner, i.e., $\text{diag}(A)$, would not have the same attractive computational properties because its computation would require handling a dense time domain square matrix, $\mathcal{N}$. The upside of the block diagonal preconditioner is precisely that it takes care of the scanning strategy-induced increase in the condition number without dealing with the complexity of the time domain processing.}.

The effect of Jacobi preconditioners onto the eigenspectrum of \mcala\  is to shift the largest eigenvalues  towards the unity, thus potentially decreasing the condition number of the preconditioned system. However, the nearly singular eigenvalues due to the noise correlations or the filtering will not in general be accounted for. 
These are common for ground based experiments and consequently the convergence of the PCG with the block-diagonal preconditioner is often found unsatisfactory. Indeed, in extreme, albeit not uncommon, cases this is manifested as a saturation of the residuals level and lack of the actual convergence down to a required threshold~\citep[e.g.,][]{2014A&A...572A..39S}.

\subsection{Two-level Preconditioners} \label{subsec:deflat}
An alternative preconditioner may be found among the class of the so called \emph{Deflation} preconditioners  that have proven to be successful in presence of few isolated extremal eigenvalues.
They act as de-projectors from the so called \emph{deflation subspace}, $\mathcal{Z}$. This subspace is   generated by $r$ linearly independent eigenvectors  related to the smallest  eigenvalues and constitute the columns of the \emph{deflation } matrix $Z$. This matrix is needed to define the projector $R$
\begin{equation}
R = \mathbb{1} - {A}Z(Z^\dag   {A}Z)^{-1} Z^\dag.
\end{equation}
The projector $R$ is $A$-orthogonal to any vector $ w \in \mathcal{Z}$ since $R{A}Z=0$. 
 In the exact precision algebra, $R$ would be a very efficient preconditioner, as for each steps of an iterative CG-like solver would be orthogonal to the null space of the $R{A}$. However, we deal with finite precision arithmetic and the zero eigenvalues are often as bothersome as the small ones due to the numerical precision of the machine.   

This issue can be solved by combining the operator $R$ with the Jacobi preconditioner  as it has been proposed in~\citet{2014A&A...572A..39S}
 \begin{align}
 \md\equiv&  \mbd R + ZE^{-1} Z^\dag   \nonumber \\
 =& \mbd(\mathbb{1} - {A}Z(Z^\dag   {A}Z)^{-1} Z^\dag  ) + ZE^{-1} Z^\dag  , \label{eq:m2l}
 \end{align}
 where $E$ is the \emph{coarse} operator, defined as $E = Z^\dag   {A}Z$. 
$\md$ is referred as the \emph{two-level} preconditioner and we note that it indeed fixes the issue of the  zero eigenvalues since they are rescaled all to one. Indeed,
\begin{equation}
\md {A}Z =Z.
\label{eq:ident}
\end{equation}

\noindent  The dimension of  the deflation subspace, given by $dim (\mathcal{Z})= r$ is by  construction much smaller than $N_p$, and it is straightforward to invert the matrix $E$.  Moreover, as \mcala\ is SPD, so is $E$. 

\noindent We can summarize the action of the $\md $ preconditioner,  applied on a vector $v$, as  a projection of the vector $v$ onto two  different  subspaces, namely $\mathcal{Z}$ and its orthogonal complement  $\mathcal{Y}$.  The components of $v$  are projected onto  $\mathcal{Z}$  via the $ZE^{-1} Z^\dag $ term in \eqref{eq:m2l}. In this subspace, the inverse \mcala\ is  very well approximated by $\md$,  since we have that \eqref{eq:ident} holds for any $z \in \mathcal{Z}$. On the other hand,  $\md$ acts onto a generic vector $y \in \mathcal{Y}$ in the same way $\mbd$ does, since
\begin{displaymath}
\md {A}y = \mbd {A}y.
\end{displaymath}

\noindent Thus, once $\md $ de-projects from the deflation subspace, it performs   the  PCG by means of the standard  preconditioner and it converges faster  since $\mbd {A}$  has a smaller condition number $\mathcal{O}(10)$ (due to the considerations made at the end of Sect.~\ref{subsec:jacobi}).  

\noindent It may appear that in order to build the deflation subspace, one would require the knowledge of the entire eigenspectrum of \mcala\ to determine the eigenvectors with the smallest eigenvalues.  However,  \citet{2014A&A...572A..39S} has proposed that approximated eigenpairs derived with the help of the so called \emph{Ritz approximations} (see Appendix~\ref{subsec:arnoldi}) is sufficient for this purpose.

\section{The case of the ground based experiments.}
\label{sec:degen}
 
 A ground-based \cmb experiment, scanning the sky with a focal plane including thousands of polarization sensitive pixels, has  to cope with both atmospheric and ground emissions, which have to be treated on the time domain level. This can be achieved by applying filtering to the data as discussed in Sect.~\ref{sec:filters}. The specific templates often applied in this context~\citep[e.g.,][]{Poletti2016} are as follows.
 
\subsection{ Atmospheric emissions and noise correlations }
\begin{figure}
\includegraphics[width=1.\columnwidth]{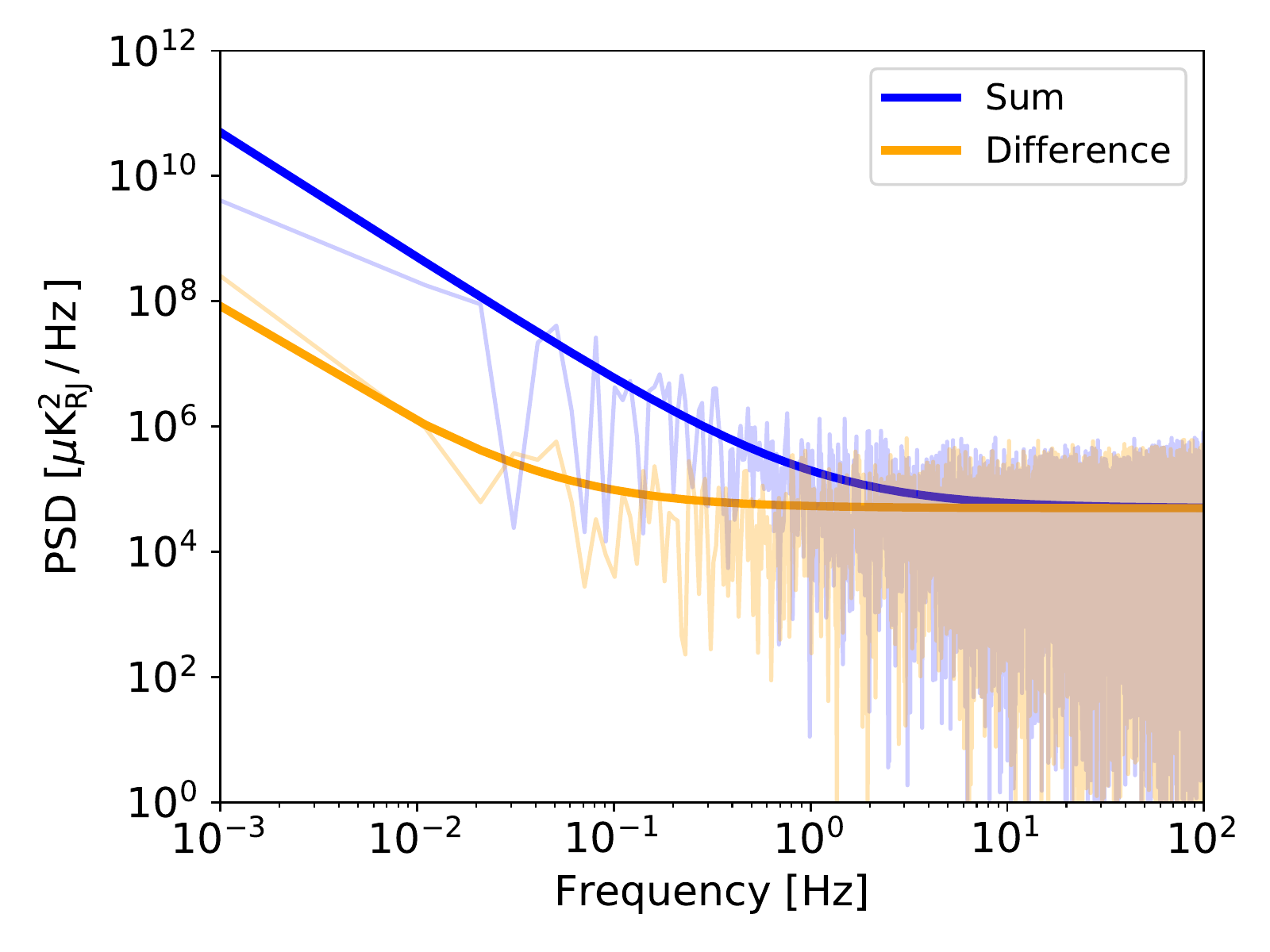}
\caption{ Power Spectral Density of summing and differencing the signal from simulated data. Notice the $f_{knee}/f$ dependence at small frequencies, and the flattening due to white noise  above $1 \si{. \hertz}$. Solid blue (orange) line refers to a signal with a $f_{knee}= 1\, (0.05) \si{. \hertz}$.}\label{fig:psd}
\end{figure}

 \noindent Both the atmosphere fluctuations, $a$, as well as detector noise, $n$, introduce contributions correlated on long time scales. While such effects could be potentially suppressed by adopting an appropriate weight matrix, $W$, in practice such a solution is prohibitive given the sizes of the current and anticipated data sets in the time domain. In such cases the diagonal weight matrices, while straightforward to operate on, will  typically lead to poor quality estimates of the sky signal with strongly correlated, spurious features appearing along the scan directions.

Such long temporal modes can be however well approximated by an arbitrary linear combination of piece-wise low order polynomials. Collecting these templates in a matrix $B$, we can express the resulting residual as,
\begin{equation}
w \equiv  a + n - Bx.
\label{eq:polyfiltering}
\end{equation}
\gpedit{Since $Bx$ is supposed to capture the correlated component of $a+n$, $w$ is expected to be approximately described by a white noise statistics. Consequently, a diagonal weight matrix can be now sufficient for obtaining good quality sky maps. In our simulations we assume this approximation to be exact and thus simulate TODs with piecewise-stationary white noise.}\\*

Filtering these particular modes results  in removing from both noise and  signal long term trends present in the TODs, whose signal-to-noise ratio is usually very low. Even though, the stripes in the reconstructed map disappear we have to remember that the constraints on the large angular scales are weak. The system matrix $A$ encodes this information: the presence of the filtering operator (see the left-hand side of \ref{eq:linearsystem}) results in low eigenvalues corresponding to long modes.

\subsection{Ground pickup}

Though ground-based experiment are designed to have very low far side lobes of the beam, the signal from the ground is not negligible compared to the \cmb one.  The elevation is typically constant during an observation period and therefore the ground signal can be considered as a function of  the azimuth. If we neglect contributions from other signals, the TOD data model can  be written as:
\begin{displaymath}
d= Ps + Gg +n.
\end{displaymath} 
 \noindent Intuitively, we can think of the second term as  the ground template map $g$ projected to the time domain by means of the ``ground-pointing matrix'' $G$. This matrix has a column for each azimuthal bin, the entries of the column are equal to 1 whenever the azimuth of the pointing direction falls within the bin range, and they are zero elsewhere. 

\subsection{Map-making for the ground based experiments.}

The effects discussed earlier in this Section have typically amplitudes significantly higher than those of the sky signals, which moreover do not average out efficiently while projected on the sky. For this reason they need to be treated explicitly in the map-making process by introducing appropriate filters.
As elaborated on in~\citet{Poletti2016}, the required filters, while dealing efficiently with the unwanted contributions, may however render the system matrix \mcala\ ill conditioned, i.e. $\kappa\gg 1 $, implying the existence of degeneracies between a certain sky signal  $\tilde{s}$ and the amplitude of a certain template $\tilde{y}$. This means that some particular mode of sky signal is impossible to reconstruct  whenever the template $\tilde{y}$ is filtered out, since $P\tilde{s} = T \tilde{y} $ and $P^{\dag}F P\tilde{s} = P^{\dag} FT \tilde{y}=0$. 

This may be particularly acute in the case of the ground pick-up filtering. For a constant-elevation one-detector observation filtering ground-stationary signal results in unconstrained modes that are constant in the right-ascension direction. \gpedit{ An extensive discussion on the properties and structure of the $F_T$ matrix can be found in \citet{Poletti2016}}.

Accumulating multiple detectors and observation periods can partially break these degeneracies, but the constraints on these modes will typically be weak.

As pointed out earlier, the presence of small-eigenvalues in the eigenspectrum of the system matrix, $A$, can significantly affect the convergence of the iterative solvers and can not be accounted for by the standard, block-diagonal preconditioner.
 
\section{The Simulated Data Set}\label{sec:sims}

In this section we describe the  experimental setup we adopted to perform map-making runs.   
\noindent We exploit the simulations capabilities of the \texttt{Systematics For CMB (S4CMB)} package\footnote{https://github.com/JulienPeloton/s4cmb} to produce simulated data sets for different experimental configurations of a ground-based experiment located in the Atacama desert in Chile at an altitude of  $\si{\num{5190}}~\si{\meter}$. We remind that the site location has implications for the properties of the observation. For example, since every pixel is observed at different elevations, the projection of the scan on the sky crosses the pixel with different direction, increasing the so-colled cross-linking (i.e. the coverage in the orientation of the attach angle).

We consider a \SI{60}{\cm}-wide focal plane hosting dual-polarization pixels  sensitive to \SI{148}{\giga\hertz} with a fractional bandwidth of $26 \%$. The resolution of the telescope is assumed to be $3.5$ arcmin.

We consider three cases that differ for the target sky area and the sensitivity of the instrument, they are summarized in table \ref{tab:specs}. 
The configurations labelled Small Patch (SP) and Large Patch (LP) refer to the characteristics of current and forthcoming \cmb experiments observing either small ($f_{sky} < 1\%$) or wide ($f_{sky} \gtrsim 1\%$) sky patches.  For both cases, the Noise Effective Temperature (NET) per detector is $\sim 500 \mu\mathrm{K \, \sqrt{s}}$  but  in LP we increase of about one order of magnitude the number of detectors in the focal plane, from $600 $ to $8,000$.

We  consider  an additional case to reproduce a next generation of ground-based \cmb experiments that will observe a wider fraction of the sky ($f_{sky} \sim 20 \%$) with an increased detector sensitivity NET$\sim 360 \mu\mathrm{K \, \sqrt{s}}$ and a larger number of detectors ($50,000 $). We refer to this setup as the Very Large Patch (VLP). 

The simulated observations are divided into constant elevation scans~(CESs) during which the telescope scans back and forth in azimuth at a speed of  $\si{\ang{0.4} \second^{-1}} $ and at constant elevation  (hereafter, we commonly refer to  each azimuthal sweep as  a \emph{subscan}).   When the patch has moved out of the field of view, the telescope moves the elevation and azimuth toward the new coordinates of the patch and a new CES starts. 
The samples are acquired at a rate  of \SI{8}{\hertz}, which given our scanning speed is sufficient to reach $\ell \sim 1200$. The number of samples per CES depends on the width of the subscan  and on the number of detectors performing the measurements. 

\begin{table*}[htpb!]
\caption{Properties of the different scanning strategies}\label{tab:specs}
\centering 
\begin{tabular}{ lccccc }
\toprule
{Case }    &  $N_t$   &$ N_p$ &$f_{sky}$&$ NET_{array} \, [\mathrm{\mu K \sqrt{s}} ] $& {Observation time [yr]}  \\
\midrule
{Small  Patch } & $\si{\num{3d10}}$	&$\si{\num{4d4}}$ &$0.1 \% $& 20.4    & 2 \\
{Large  Patch } & $\si{\num{3d10}}$  & $\si{\num{2d6}}$& $5 \% $ & 5.6 &1 \\
{Very  Large Patch } &$ \si{\num{3d10}}$ &$ \si{\num{1d7}}$ &$ 20\%$ & 1.6 &1 \\
\bottomrule
\end{tabular}
\end{table*}
 
Using the simulated scanning strategies we scan an input \cmb map computed with the \texttt{synfast} routine of  the Hierarchical Equal Area Latitute Pixelization  (HEALP{\scriptsize IX} library, \citet{2005ApJ...622..759G}) \footnote{http://healpix.sourceforge.net} and then add a white noise realization corresponding to the sensitivity of each experimental configuration. The input signal power spectrum has been computed with the CAMB package \citep{Lewis:1999bs} assuming the \pl\ 2015 best fit cosmological parameters \citep{2016A&A...594A..13P} and $r=0.1$. Since we consider maps at the resolution of $3.5$ arcmin, we sampled the input CMB signal on a grid with $1.7$ arcmin resolution  (corresponding to a resolution parameter $\texttt{nside}=2048$).
The definition of the observed pixels is performed prior to the map-making procedure and it is based on discarding those samples which do not observed pixels with enough redundancy. 

\subsection{Data model} \label{subsec:datamodel}
The simulated TODs acquired by each detectors can be expressed as in Eq.~\eqref{eq:datamodel}. However, since the detectors are grouped in pairs sensitive to orthogonal polarization states of the light, commonly referred as $d_{top}$ and $d_{bottom}$, the signal coming from the two  detectors of a given pair can be combined in order to disentangle the total intensity and polarization signals without any loss of accuracy by summing and differencing two signal:
\begin{displaymath}
d^{\pm}\equiv\frac{1}{2}\left( d_{top} \pm d_{bottom}\right).
\end{displaymath}

Hence, one can independently estimate intensity and polarization (expressed via the $Q,U$ Stokes parameters) maps. Two separated data models can be written for the signal and noise component of the time streams:
\begin{align}
d^+_t \equiv&P_{tp}I_p +  n_t^+, \label{eq:d+}\\
d_t^- \equiv &P_{tp} ( \cos (2\phi_t) Q_p  + \sin(2\phi_t) U_p ) + n_t^-\label{eq:d-}
\end{align}
where $n^{\pm} $ is the noise term and can be analogously defined as $d^{\pm}$.  \gpedit{ We note that in the following we will neglect possible systematic effects that can leave a residual intensity component in $d_t^-$, such as a calibration mismatch between the two detectors of a pair. The noise properties of the sum and difference time streams are different, as depicted in Fig.~ \ref{fig:psd}, }and therefore require a different set of low order polynomials to be filtered out. In our analysis, the time stream $d^{-}$ if filtered  by  zeroth and first order polynomials, $d^+$ by the first three order polynomials  and we assume that these filters completely remove the $1/f$ component. For this reason our noise simulation contain solely white noise. Simulating the correlated noise component is important when evaluating the end-to-end performances of an experiment, but in this paper we focus only on the performances of the map-making solver, which depend mostly on the scanning strategy and data processing adopted.
The sum-difference approach and the fact that $n_t^+$ and $n_t^-$ are uncorrelated allow to separate the intensity and the polarization reconstruction, we take advantage of this by focusing only on the latter for the rest of this paper.

The ground template is the same for summed and differenced data, and its column number is the same as the number of azimuthal bins ($100\div 1000$, depending on the width of the patch). Each azimuth bin has a fixed width of $\si{\ang{0.08}}$. The rows are as many as the number of samples in each CES $N^{CES}_t$.

 For simplicity, in the following analysis, we do not build $F_{[G,B]}$, instead, we avoid the burden of explicit orthogonalization of the filters by using as the filter $F_T$ in Eq.~(\ref{eq:linearsystem}), a simplified filter given by $F_B W^{-1} F_G W^{-1} F_B$, which is explicitly symmetric and would have been equivalent to $F_{[G,B]}$, were the filters, $F_G$ and $F_B$, be orthonormal from the outset~\citep{Poletti2016}.

\section{Constructing two-level preconditoner}

\label{sec:constr2lvl}

\begin{figure}[h]
\includegraphics[width=1\columnwidth]{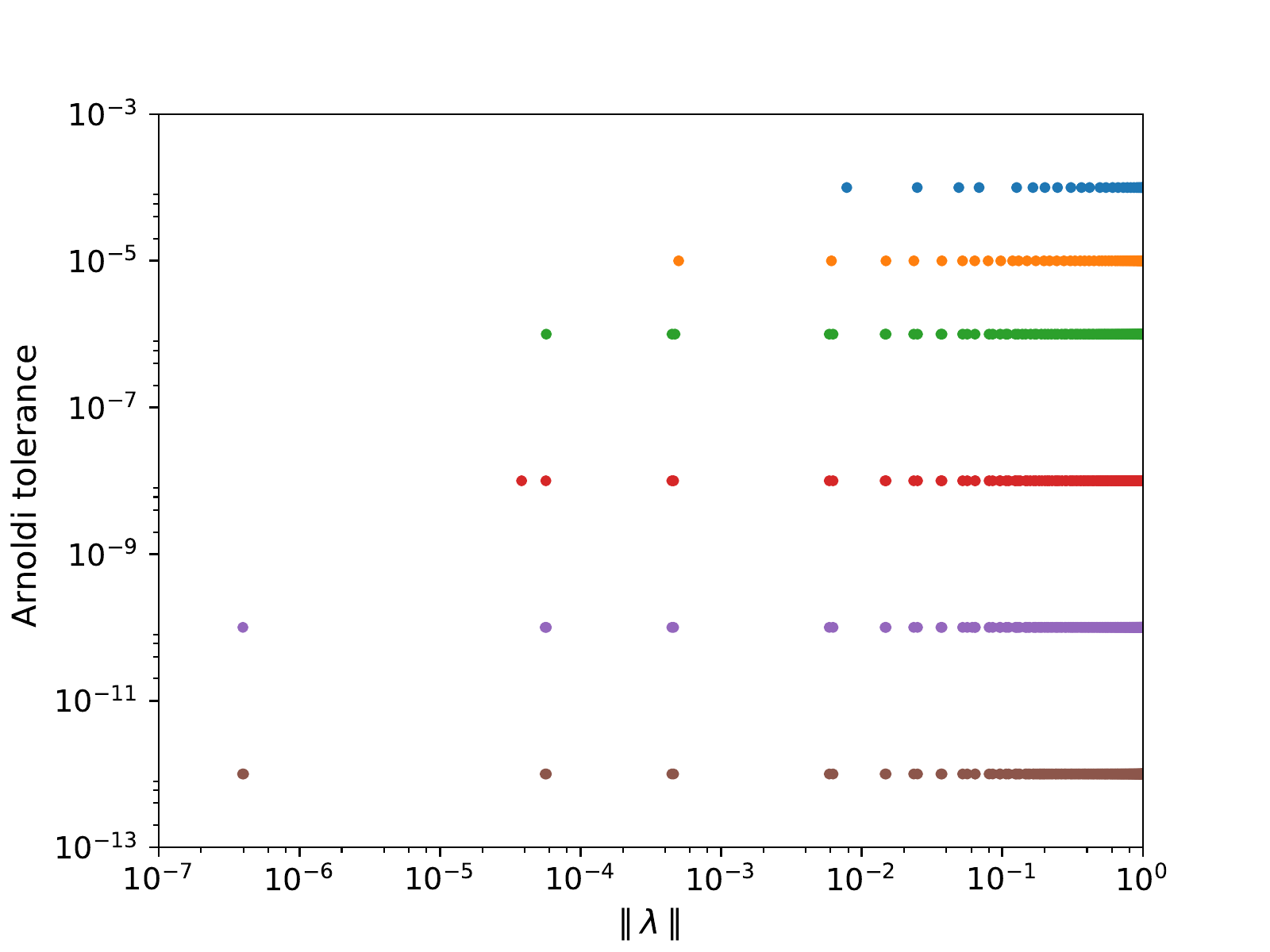}
\caption{Ritz Eigenvalues $\mbd A$ estimated with different Arnoldi algorithm tolerance , $\epsilon_{Arn}$. An histogram version of this plot can be found in Fig.~\ref{fig:binned_ritz}.}\label{fig:eigenspectrum}
\end{figure}

A construction of the two-level preconditioner requires knowledge of the deflation operator, ${\cal Z}$. We estimate it following the prescription of~\citet{2014A&A...572A..39S}, which employs the Arnoldi algorithm to compute approximate eigenpairs of the matrix $B=\mbd {A}$. A suitable selection of these is then used to define the deflation operator, $Z$. This process has two free parameters, $\epsilon_{Arn}$ and  $dim(\mathcal{Z})$, that we discuss in the rest of this section and fix them in Sect.~\ref{sec:results} using numerical experiments.

The Arnoldi algorithm iteratively refines the approximate eigenpairs of the provided matrix and ends the computation when a given tolerance, $\epsilon_{Arn}$, is reached (see Appendix~\ref{subsec:arnoldi} for more details). The lower the Arnoldi tolerance, the larger is the rank of the approximated $B$ and, consequently, the larger is the number and the accuracy of the estimated eigenpairs. In Fig.~\ref{fig:eigenspectrum}, the approximated eigenvalues are reported for several choices of $\epsilon_{Arn}$ and some specific choice of the system matrix corresponding to the small patch case discussed later. It shows that not only the number but also the range of the eigenvalues increase with smaller tolerance. This is intuitively expected since the Arnoldi algorithm relies on the \emph{power method}~\citep[e.g.,][]{Golub:1996:MC:248979}, and thus it first estimates the largest eigenvalues before moving to the smaller ones. Once the tolerance is as small as $\num{d-9}$ the range of the estimated eigenvalues starts to saturate. If we attempt to reach a threshold smaller than $\sim$\num{d-12}, the Arnoldi iteration proposes a new search direction that, due to the finite numerical precision, has no component linearly independent from the previous ones. Consequently, from that moment on the algorithm keeps producing eigenvalues equal to zero that are not eigenvalues of $B$ but just the sign that the Arnoldi algorithm has converged and exhausted its predictive power. In the studied cases we find that this typically requires  $\sim 150$ iterations.

The computational time required for each Arnoldi iteration is similar to the CG but the memory consumption can be very different: while the  Arnoldi needs space for as many vectors as the iteration number, the CG requires only few vectors in the memory regardless of the iteration number. However, this is not a problem if the size of the map is negligible compared to the time streams -- a condition likely to be met in forthcoming ground-based experiments.

The other parameter in the construction of the preconditioner is the dimension of the deflation space. For any given Arnoldi tolerance, this can be either fixed directly by defining the number of the smallest eigenvalue and eigenvectors retained to construct ${\cal Z}$ or by defining a threshold below which the eigenvalues and eigenvectors are retained, $\epsilon_{\lambda}$.

\begin{figure*}[htpb!]
\includegraphics[width=2\columnwidth]{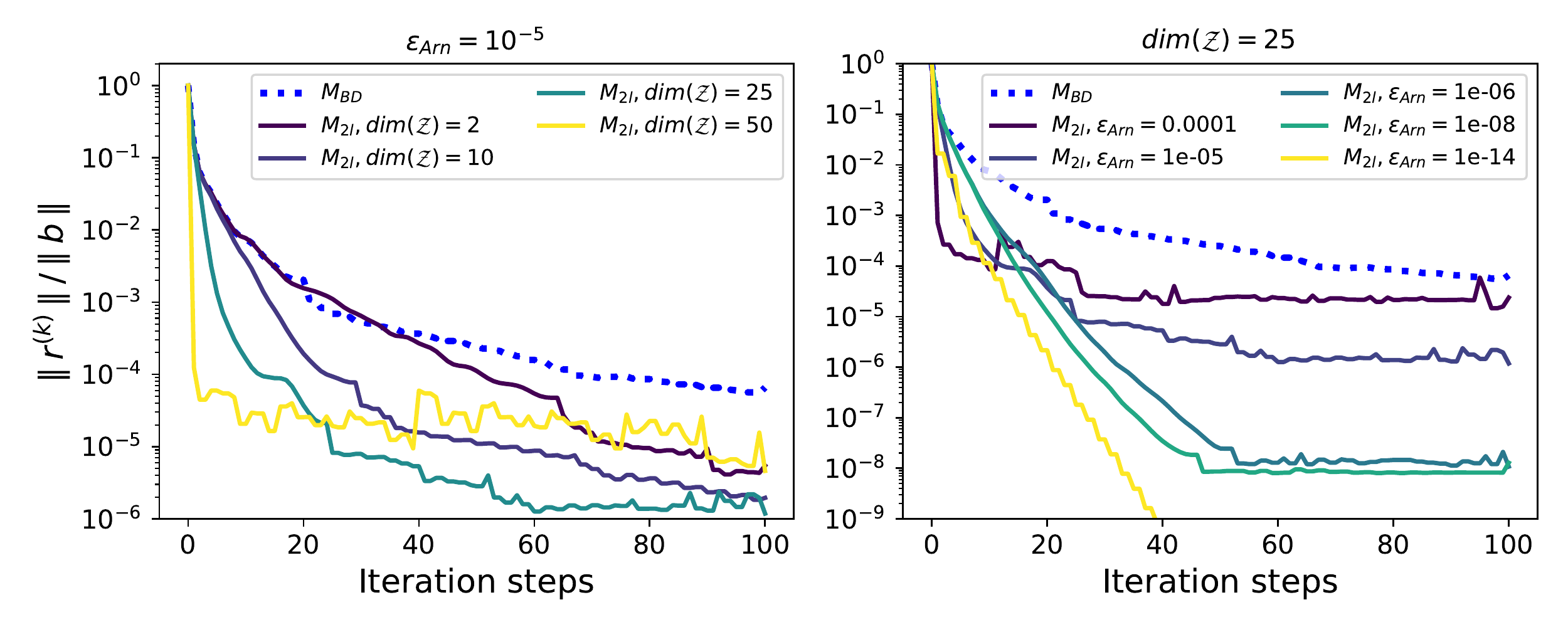}
\caption{ (left)  PCG residuals  with $\md$ and $\mbd$ preconditioners for several choices of the size of the deflation subspace, Ritz eigenpairs are computed up to a fixed Arnoldi tolerance $\epsilon_{Arn}$. (right) The $\md$  is built by selecting  a fixed  number of Ritz  eigenvectors, \ie\  $dim (\mathcal{Z})=25$, computed by running the Arnoldi algorithm with several choices for $\epsilon_{Arn}$.}\label{fig:arntol_vs_dimZ}
\end{figure*}

\section{Results and discussion}\label{sec:results}

In this section, we present the performance comparisons of the standard block-diagonal and the two-level preconditioners both applied on simulated noise or signal-only dataset observing with the scanning strategies listed in Table \ref{tab:specs}. Moreover, we focus on the reconstruction of the polarization component of the sky,  but the results for total intensity are similar and are  reported in Appendix~\ref{app:Truns}.\\*

\subsection{Comparison methodology}

We use three types of metric in order to estimate the level of accuracy achievable by each considered approach. First we consider the norm of the standard map-level residuals  \gpedit{as defined  in Eq.~\eqref{eq:residuals} normalized by the norm of the right hand side, \ie\ } 
\begin{equation}
\parallel \hat{r}^{(k)} \parallel \equiv \frac{\parallel r^{(k)} \parallel }{\parallel b\parallel }.
\end{equation}

This measure of convergence is naturally provided in the CG algorithm and, most important, does not require the knowledge of the true solution. It is indeed the one typically employed in real applications for measuring the reconstruction quality.

In order to get further insights, in this paper we also consider metrics that require knowledge of the exact solution, which is available only in the case of signal-only simulations.
We make use of the norm of difference between the true and recovered map, \gpedit{ defined in Eq.~\eqref{eq:errnorm}}, and the \emph{bin-by-bin} difference between the power spectrum of the input map and reconstructed map, binned using equally spaced bins in multipoles, $\ell_b$, 
\begin{equation}
\Delta \mathcal{C}_{\ell_b}^X \equiv \frac{|\mathcal{C}_{\ell_b}^{X,in} - \mathcal{C}_{\ell_b}^{X,out} |}{\sigma^X_{CV}} , \label{eq:deltapowsp}
\end{equation}
with $X= E, B $;  the differences are normalized with respect to the   \emph{cosmic variance}  of the input \cmb map,
\begin{equation}
\sigma^X_{CV}(\ell_b,\mathcal{C}^X_{\ell_b} ) \equiv \sqrt{\frac{2}{(2 \ell_b +1) f_{sky } \Delta \ell_b}} \mathcal{C}^X_{\ell_b}.
\end{equation}
This power spectrum difference enables to check which scale in the maps are better constrained, and the normalization gives an estimate of how much the signal \emph{intrinsically } fluctuates. We stress that we compare against the power spectrum of the input map, not the power spectrum used to simulate it. Therefore, the normalization is just a reference value and $\Delta \mathcal{C}_{\ell_b}^X$ has no cosmic variance.\\*
As the considered sky patches cover only a fraction of the sky, the power spectra are computed using a pure-pseudo power spectrum estimator \xpure\ \citep{2009PhRvD..79l3515G}.  This is a pseudo power spectrum method \citep{2002ApJ...567....2H} which corrects the \emph{E-to-B-modes leakage } arising in presence of incomplete sky coverage \citep{PhysRevD.76.043001,2003PhRvD..67b3501B, PhysRevD.65.023505}.

\subsection{Setting the two-level preconditioner}
\label{sec:twoLevelParameters}
We use numerical experiments to show the role of the two free parameters involved in the computation of the two-level preconditioner, $\epsilon_{Arn}$ and ${\rm dim}\,{\cal Z}$.
A sample of the results is shown in Fig.~\ref{fig:arntol_vs_dimZ}. In the left panel we fixed the Arnoldi tolerance to $\epsilon_{Arn}= 10^{-5}$ and change the dimension of the deflation space from ${\rm dim}\,{\cal Z} = 2$ to $50$, which corresponds to varying $\epsilon_\lambda$ from $0.01$ up to $1$. 

The size of the deflation subspace affects strongly the steepness of the initial convergence. This is expected because, if we use all the vectors produced by the Arnoldi algorithm to construct the deflation subspace, the residual after the first iteration is related by construction to the Arnoldi tolerance. On the contrary, the case with the block-diagonal preconditioner corresponds to ${\rm dim}\,{\cal Z} = 0$. The more we include vectors in the deflation subspace, the more we approach ${\rm dim}\,{\cal Z} = 50$, which retains nearly all the vector produced by the Arnoldi iterations and indeed jumps immediately to a residual close to the Arnoldi tolerance. 
\gpedit{In terms of the final residual, it seems that the larger ${\rm dim}\,{\cal Z}$ the better the performances. However, the gain might become limited as all the cases adopting the two-level preconditioner saturate on a similar level of residual $\|\hat{r}^{(k)}\| \sim 10^{-5} \div 10^{-6}$ . Moreover, the case with the largest basis, ${\rm dim}\,{\cal Z} = 50$, seems to enter a regime of high numerical noise that causes it to deliver worse final convergence compared to the other cases. Therefore, this numerical experiment suggests that, for fixed Arnoldi threshold, it is not recommended to strive for the largest deflation basis because an intermediate value can yield comparable (or even better) final convergence and consume less memory and CPU time -- both proportional to ${\rm dim}\,{\cal Z}$.}
In our set up, the case with $\epsilon_\lambda = 0.2$ (corresponding to ${\rm dim}\,{\cal Z} = 25$) delivers slightly more accurate estimate and will be our value of choice in the rest of this paper. As the threshold of $10^{-6}$ is commonly adopted in the CMB map-making procedures for the convergence, these residuals are already quite satisfactory. Moreover, they are also already nearly two orders of magnitude better than what can be achieved with the standard, block-diagonal preconditioner.

We would like to make sure that better precision could be reached if needed.  In the right panel of the figure we fix ${\rm dim}\,\mathcal{Z}$ to $25$ and show how the performances change as the  Arnoldi tolerance threshold decreases. The more we decrease the Arnoldi threshold, the lower value we get for the final residuals -- for the reason discussed above, the first few tens of iterations are affected by the the fraction of eigenvectors retained rather than $\epsilon_{Arn}$ itself. Choosing $\epsilon_{Arn} \sim 10^{-6}$ seems already sufficient as it allows reaching residuals level as low as $10^{-8}$, but even lower residuals are reached by decreasing further  $\epsilon_{Arn}$. In particular, we did not reach any saturation when we let the Arnoldi converge completely, i.e. when $\epsilon_{Arn} = 10^{-14}$. We might be tempted to always use such a low threshold to build the preconditioner for our CG solver. However, when we push the Arnoldi to a given threshold we are basically solving the system to the same residual threshold with the GMRES algorithm. Therefore, building a two-level preconditioner for a given system using a value of $\epsilon_{Arn}$ much lower than the target CG residual is not meaningful. 

We thus conclude that, in order to achieve a very accurate solution  (PCG residual tolerance $\sim 10^{-7}$ or better) by means of the two-level preconditioner, the Arnoldi algorithm has to converge within a tolerance of $\epsilon_{Arn}< 10^{-6}$, and ${\rm dim}\,{\cal Z}= 20\div 30$ eigenvectors are required to build the deflation subspace.

\subsection{Exploitation of a precomputed two-level preconditioner}
 In the previous section, we have shown that building a two-level preconditioner with a fully converged Arnoldi algorithm gives the best CG convergence rate. Building such a preconditioner may not always be desirable for a single map-making run, given the extra numerical cost. Nonetheless, in this section we show that it typically not only leads to significant performance gains when many similar map-making runs are to be performed, but in process yields often better solutions for some single runs.

\subsubsection{Divide-and-Conquer map-making of one season of observation}
\label{sec:divideConquerResults}
We now explore a different scenario, the so-called \emph{divide-and-conquer} map-making, in which we solve for many map-making problems with a system matrix $A$ and right hand side (RHS)  $b$ that are similar but not equal.
CMB experiments can get the best possible map out of their observation only if they analyze the full data set at once. Nevertheless, splitting the full data volume into smaller groups and producing their maps independently can enormously reduce the computational complexity of map-making permitting to capitalize on the embarrassingly parallel character of this approach, while still producing high quality maps.

\begin{table}[htpb!]
\centering 
\small
\begin{tabular}{ l c cc cc c}
\toprule
 &   &  median   &  percentile & \texttt{kcpuh   } \\
\midrule
\multirow{ 4}{*}{SP} &$\mbd $ & $ 3\cdot 10^{-5}$ & 0.0001& $8.1^{\dag}$ \\
&$\md, \epsilon_{Arn}=10^{-6}$  &$9\cdot 10^{-6}$ &$3 \cdot 10^{-5}$  & $12.6^{\dag}$ \\
&$\md  $ simpl.  & $7\cdot 10^{-7}$&$2\cdot 10^{-6}$  & $10^{\dag}$\\
&$\md,  \epsilon_{Arn}=10^{-12}$ & $1\cdot 10^{-7}$&$3\cdot 10^{-6} $ & $20.8^{\dag}$ \\
\midrule
\multirow{ 2}{*}{LP} &$\mbd $ & $\si{\num{0.001}}$ & 0.0004	 &$10.8$ \\
&$\md  $ simpl. 		& $3\cdot 10^{-7}$&$1\cdot 10^{-7}$ & $8.5$ \\
\midrule
\multirow{ 2}{*}{VLP} &$\mbd $ & $10^{-5}$ & $ 10^{-5}	 $  & $19.2$\\
&$\md  $ simpl. 		 & $ 10^{-7} $&$ 10^{-7}$  & $15.9$\\
\bottomrule
\end{tabular}
\caption{Median, $1\sigma$ statistics  of residual norms, $\parallel\hat{ r} ^{(k)} \parallel $ and computational cost of PCG runs for  different scanning strategies.We consider $p_{16th}$ and $p_{84th}$ respectively the  16-th and 84-th percentiles as $1\sigma$ upper and   lower bounds.  In the fourth column we quote $(p_{84th}-p_{16th})/2$. $^{\dag}$Values rescaled from \textsc{Edison}  to \textsc{Cori } computational system  to a better compare performances. }\label{tab:stats}
\end{table}

\begin{figure*}[!htpb]
\includegraphics[width=2\columnwidth]{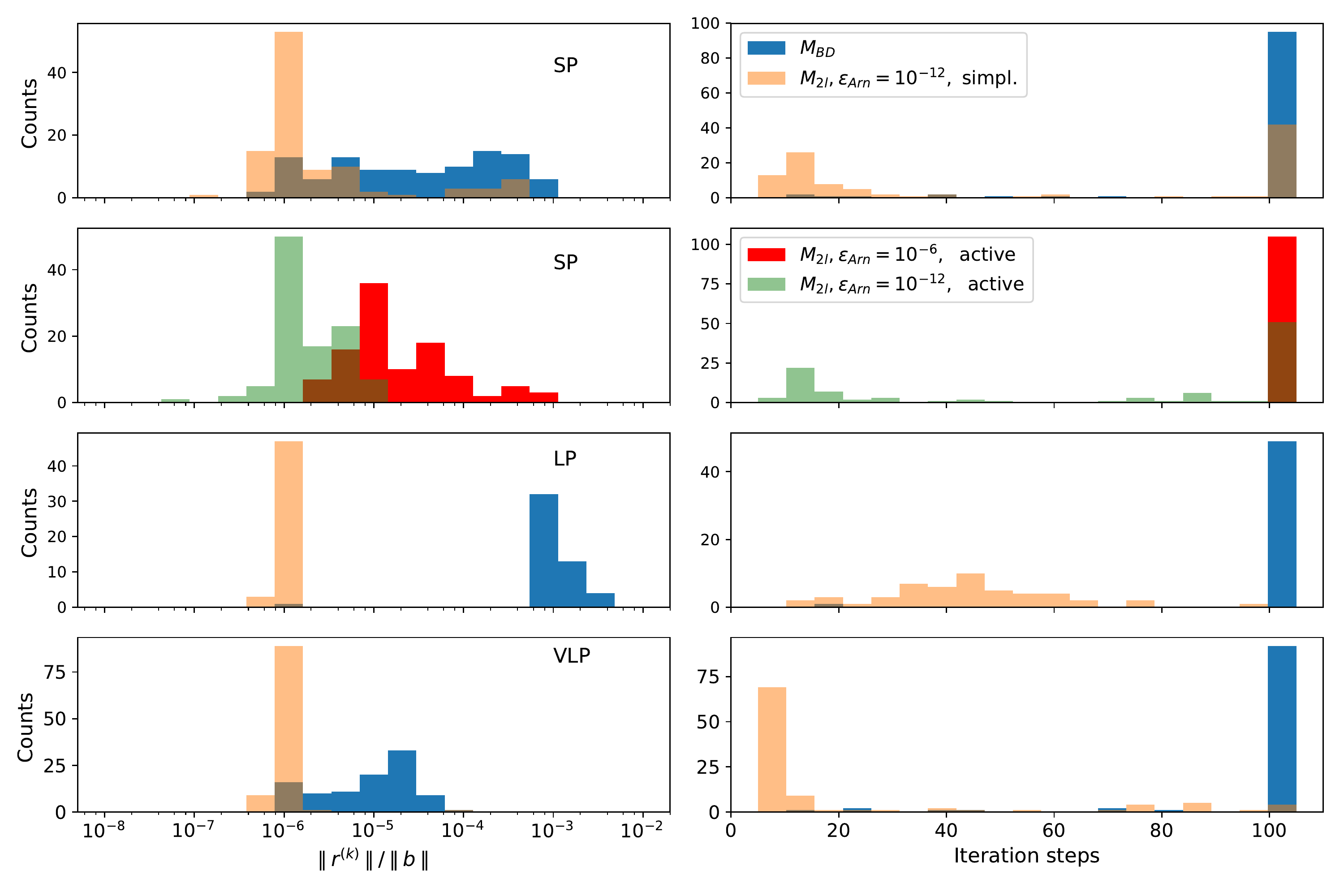}
\caption{ Histograms of (left) residual norms and (right) iteration steps of PCG runs performed on simulated data on SP, LP and VLP respectively in top, center, bottom panel. Shown with  blue bars are the histogram related to PCG runs with $\mbd$, orange bars runs with $\md$  applied with the simplified approach.  In top panel are also shown PCG runs applied with the active approach with $\epsilon_{Arn}=10^{-6} \,(\,10^{-12}\, ) $ as red (green) bars. }\label{fig:histos2D}
\end{figure*}

In the context of the ground based experiments which typically scan the same sky area repetitively multiple times, these smaller map-making problems can be defined in such way that their system matrices $A$ have similar properties.\\*
We explore the performances of the two-level preconditioner in this context starting from simulation of a two season data set of SP. 
For this scanning strategy  each CES lasts about 15 minutes and  we split the whole observation into 250 subsets consisting of 27 CESs. This subgroup roughly corresponds to all the data taken in a given day.  Each processing element performs a PCG run on one of such subsets, which is characterized by $N_t\sim \si{\num{d8}}$ and $N_p \sim \si{\num{4d4}} $. Given these numbers, we can perform as many as two PCG runs per node of the \textsc{Edison} computing system\footnote{\url{http://www.nersc.gov/users/computational-systems/edison}}, which provides 64 GB of memory.

We consider different types of two-level preconditioner runs
\begin{enumerate}
\item \textbf{The ``Active'' approach}: the Ritz eigenpairs are computed for each subset of data. We use ${\rm dim}\,{\cal Z} = 25$ ($\epsilon_\lambda = 0.2$) but consider two values for $\epsilon_{Arn}$, $10^{-6}$ and $10^{-12}$. The former corresponds to the prescription we have given in Sect. \ref{sec:twoLevelParameters} for the single map-making run. The latter corresponds to the best preconditioner we can have with this technique. We explained earlier that it is not meaningful to build such a preconditioner for a PCG run, but it provides a useful limit case to compare against.
\item \textbf{The ``Simplified'' approach}: the Ritz eigenpairs are computed only for one subgroup, using ${\rm dim}\,{\cal Z} = 25$ and $\epsilon_{Anr} = 10^{-12}$. The $\md $ built from this eigenvector basis is then applied to the rest of the whole data set. This approach is computationally cheaper than the active one, but it is less specific and it can work only in the case where the computed  deflation basis is very  representative of the whole dataset.
\end{enumerate}

\begin{figure*}[h]
\subcaptionbox{}{\includegraphics[width =.72\columnwidth]{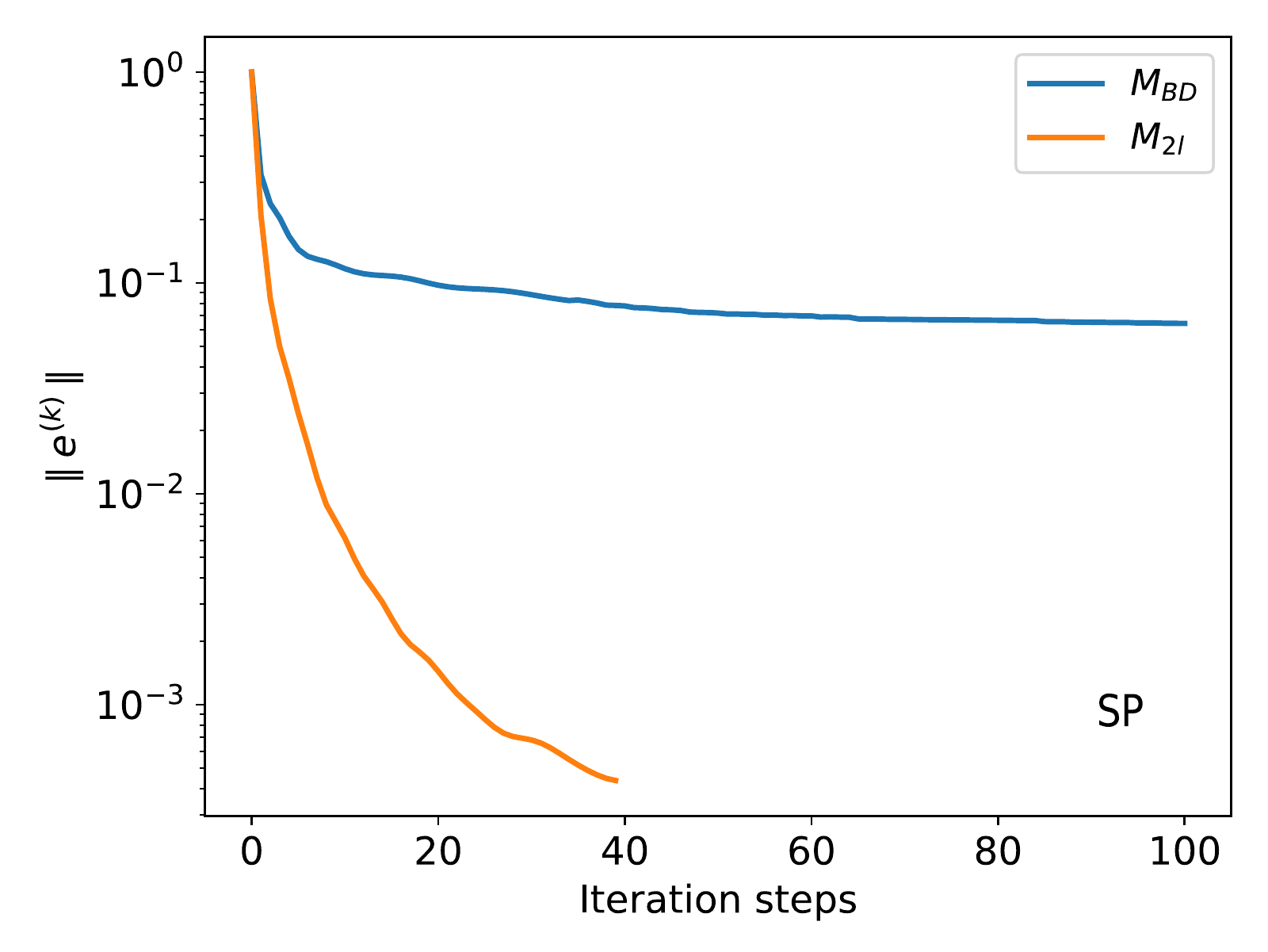}}
\subcaptionbox{}{\includegraphics[width =.68\columnwidth,clip=true,trim =1cm 0 0 0 ]{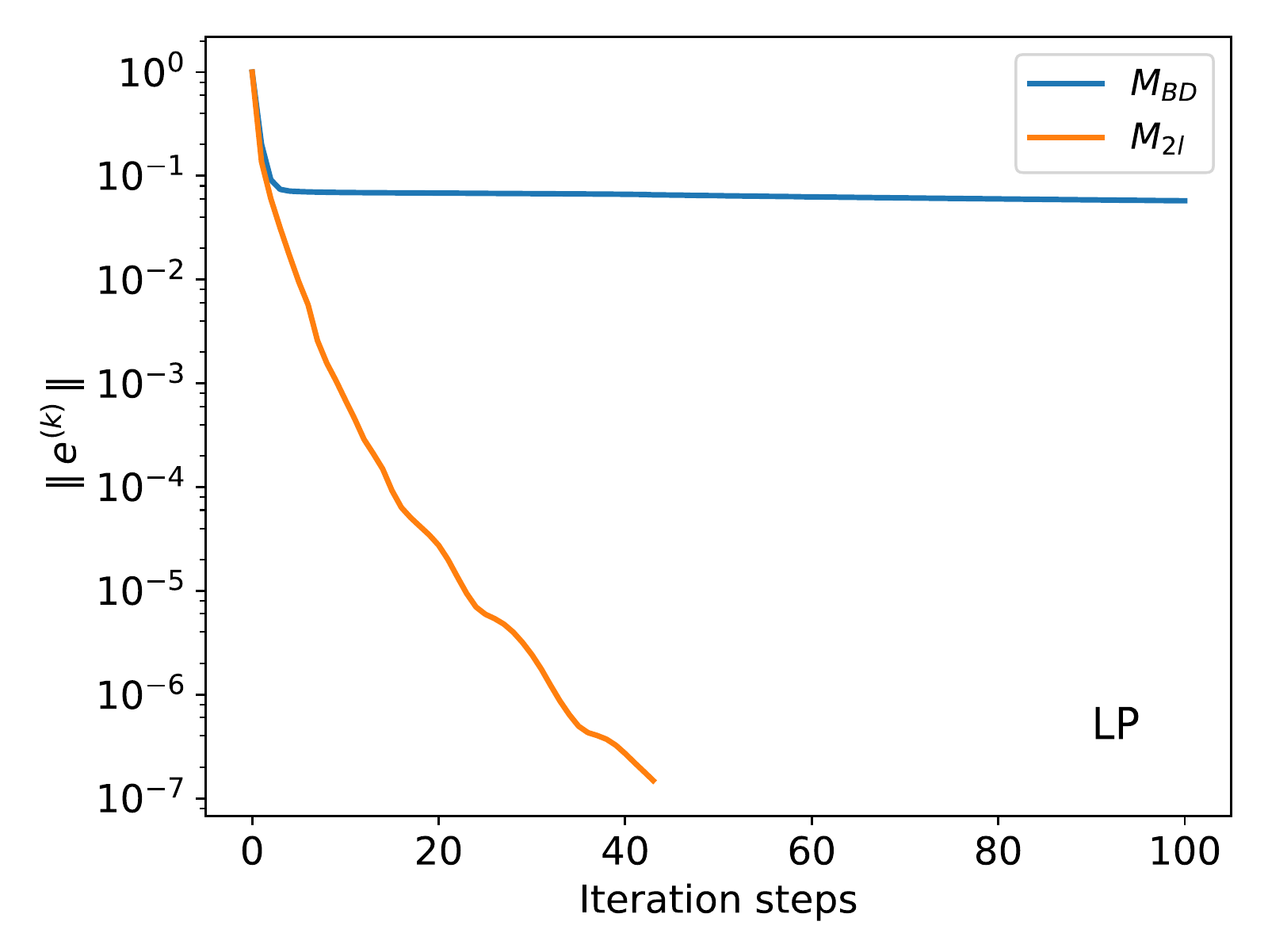}}
\subcaptionbox{}{\includegraphics[width =.67\columnwidth,clip=true,trim =1cm 0 0 0
\columnwidth]{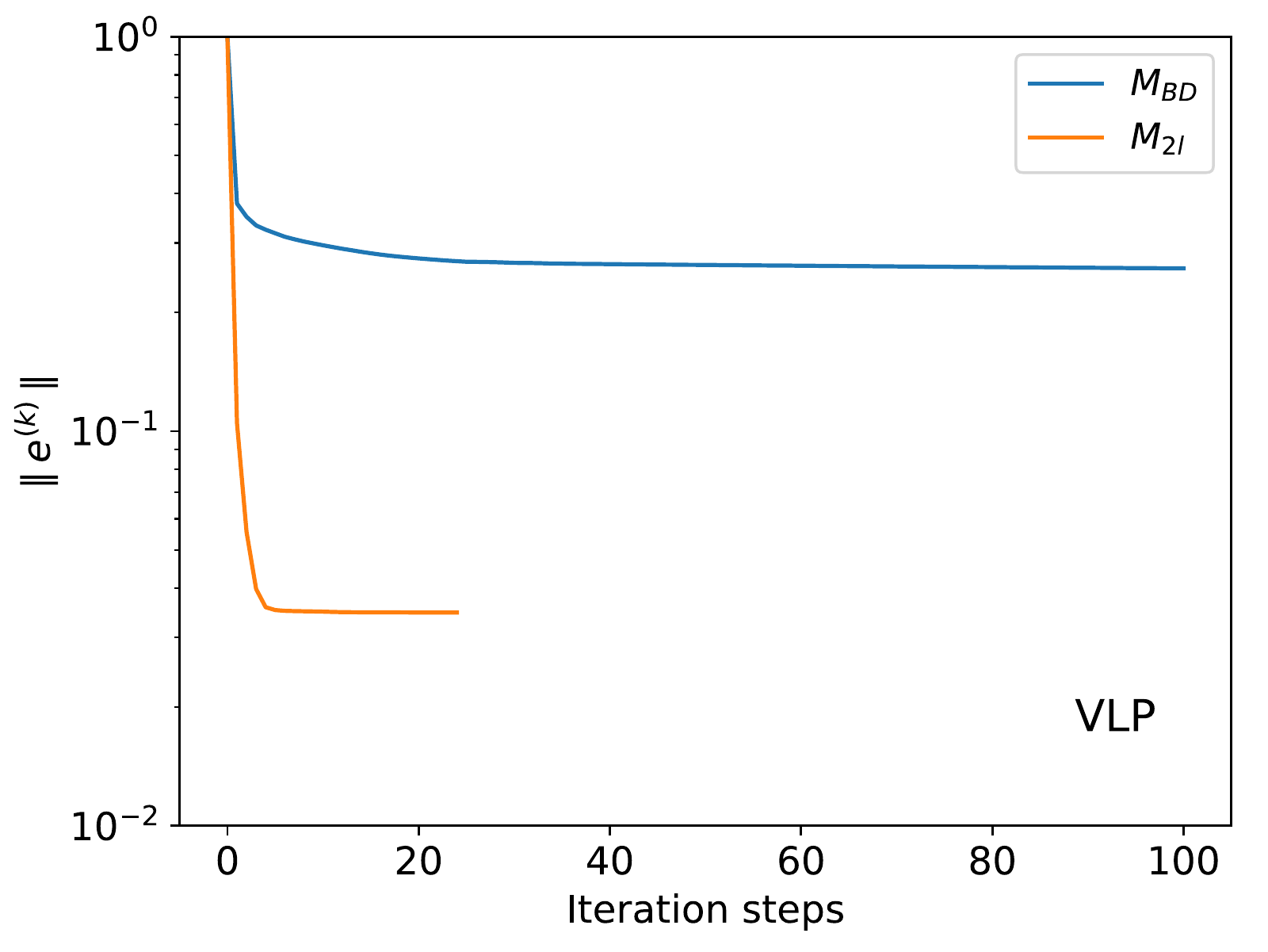}}
\caption{PCG error norms, $e^{\left(k\right)}$, as defined in Eq.~\eqref{eq:errnorm},   for a group of 27 signal-only CESs observing SP (left), for a group of 7 signal-only CESs observing LP (center), for a group of 3 signal-only CESs observing VLP (right). }\label{fig:resid_errs}
\end{figure*} 

Fig.~\ref{fig:histos2D} reports the histograms of the PCG performances, the left and right panel respectively show the residual at the last iteration and the total number of iterations performed, the rightmost bin of the latter collects the runs that did not meet the convergence criterion of $10^{-6}$ within 100 iterations. \gpedit{As expected,  the degeneracies in the system matrix play a role in the runs involving $\mbd$ (blue bars), so that only $\sim 30\%$ converges below a threshold of $10^{-6}$ and most of the runs converges at higher residual tolerances. As far as $\md$ is concerned, the most remarkable result is that the simplified approach (orange histogram) perform considerably better than the active case with $\epsilon_{Arn} = 10^{-6}$ (red bars) and nearly as good as the one with  $\epsilon_{Arn} = 10^{-12}$ (green bars).} This means that, rather than using a preconditioner tailored on the subset of the PCG run, it is important characterize well the most degenerate modes, even on a slightly different system: the deflation basis  computed from a subset of data is very well representative  for the  whole season data.
This demonstrates the simplified approach provides a very effective way of implementing the divide-and-conquer map-making run in the context of the ground-based observations.\\*

\begin{figure*}[!htpb]
\includegraphics[width=2.02\columnwidth]{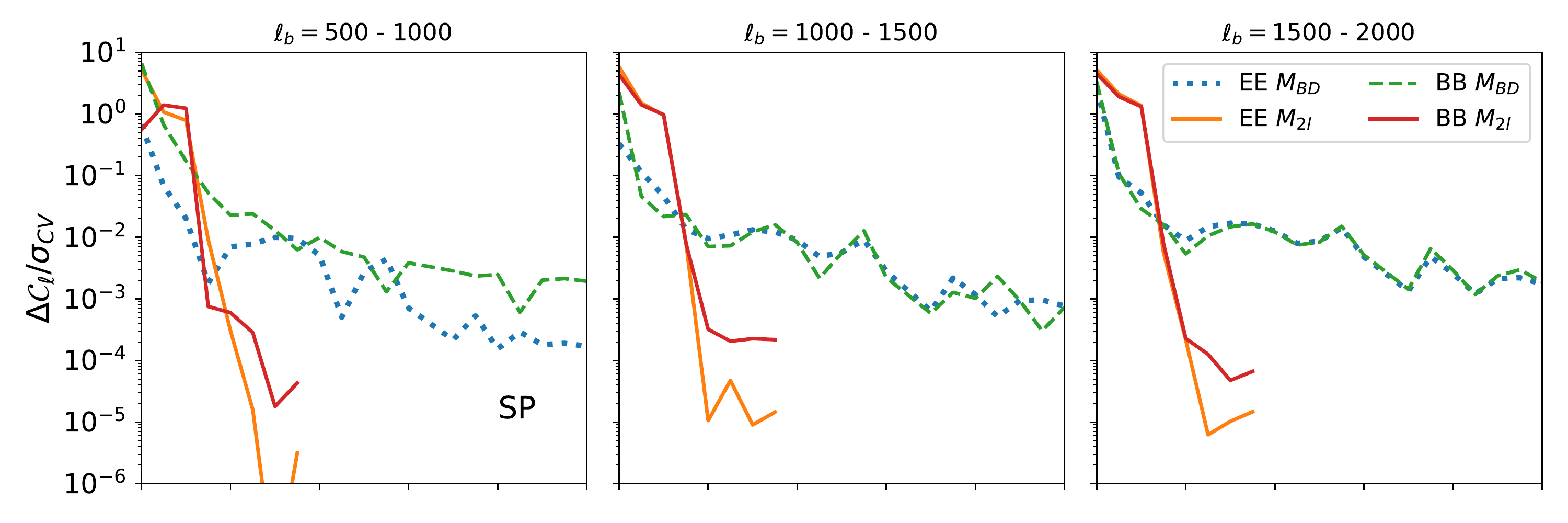}\\
\includegraphics[width=2.02\columnwidth,]{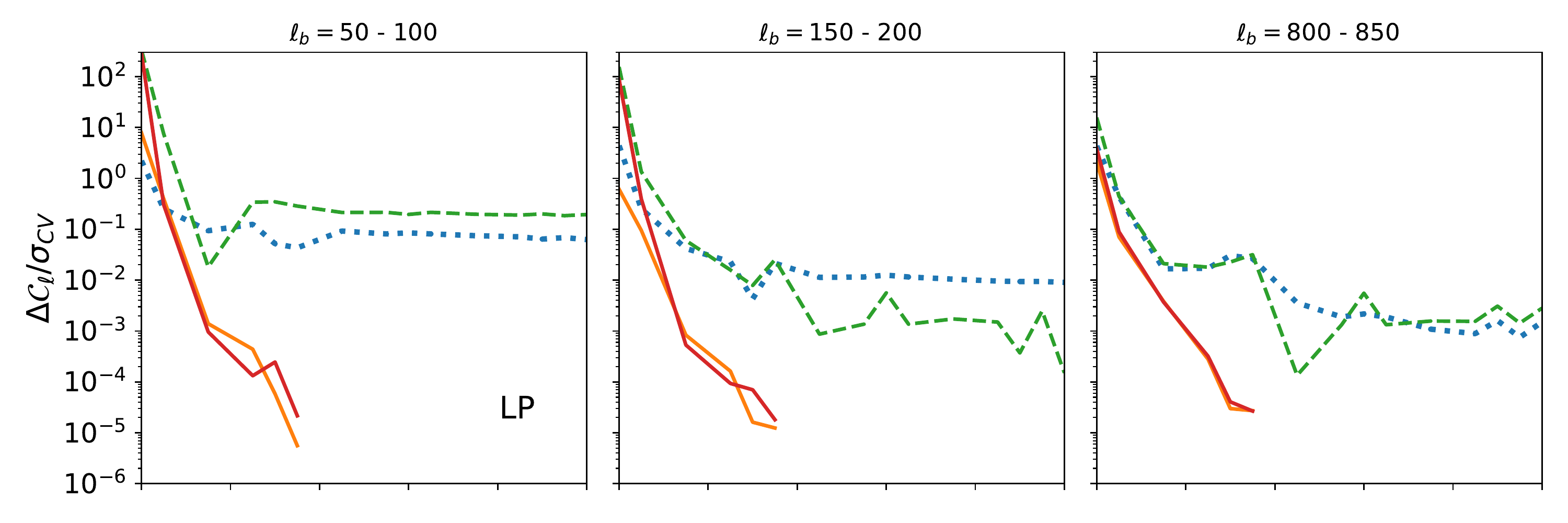}\\
\includegraphics[width=2.02\columnwidth]{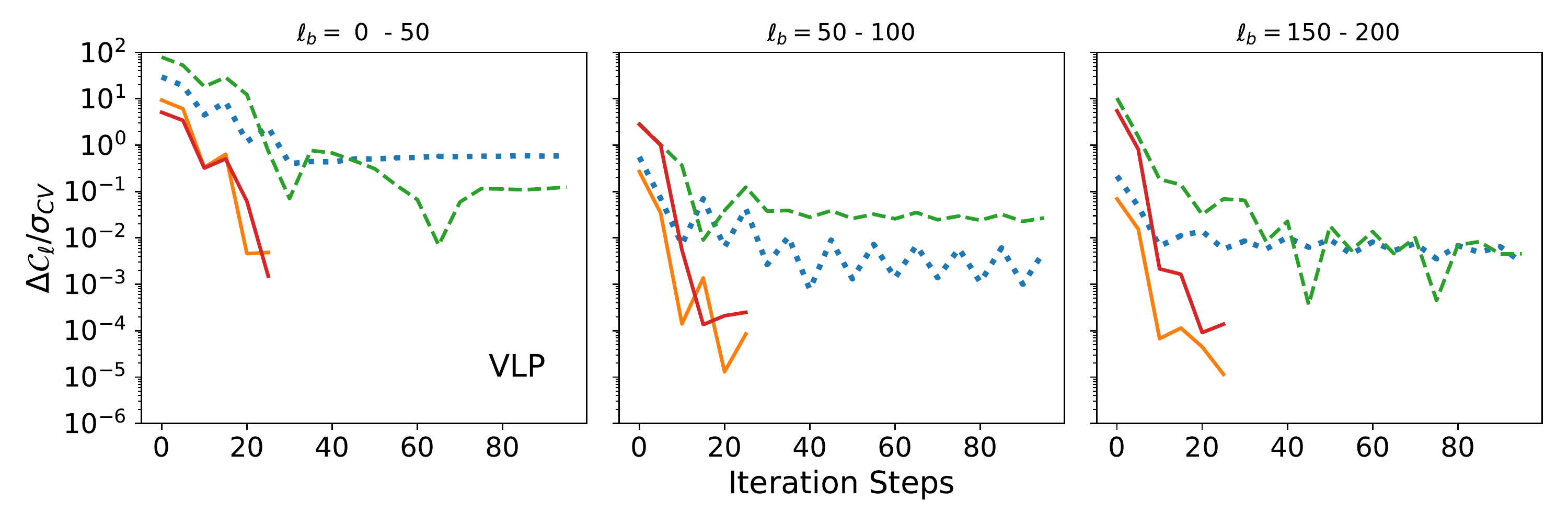}
\caption{Bin-by-bin comparison of power spectra differences defined in Eq.~\eqref{eq:deltapowsp} computed  from maps estimated at several iteration steps observing SP (top),  LP (center) and VLP (bottom). \gpedit{Dotted-blue and dashed-green lines refer respectively to  $E-$ and $B-$modes computed from maps reconstructed with Jacobi preconditioner.  Solid orange and red lines refer respectively to $E-$ and $B-$modes computed from maps reconstructed with $\md$.} We choose different multipole bins to emphasize the convergence behavior of large, intermediate and small angular scales.} \label{fig:spectrabins}
\end{figure*}

We further test this approach by applying it to observations covering a larger fraction of sky, the LP and VLP scanning strategies.  As summarized in table \ref{tab:specs} the noise level in the LP case  is about 4 times lower than the SP one: our aim is to probe the performances of this methodology in the perspective of the sensitivities that will be achieved by the forthcoming ground based experiments. For LP the length of one CES is larger than SP, each one lasting for 4 hours, and usually we simulate one or two CESs per day, depending on the seasonal availability of the patch above the horizon.  We obtain a data set consisting of 350 CESs and we chunk it into 50 subgroups made of 7 CESs.
Given the memory size of NERSC computing system  \textsc{Cori} ($128$ GB)\footnote{\url{http://www.nersc.gov/users/computational-systems/cori}}  we can run one chunk of data  made by 7 CES per node so that we distribute the seasonal data set across 50 nodes.
We construct the two-level preconditioner by taking one of these subsets and running the  Arnoldi iterations up to the numerical convergence, \ie\ $\epsilon_{Arn}=10^{-12}$. We retain the Ritz eigenvectors related to the eigenvalues smaller than $0.2$. This yields a deflation subspace with size $dim(\mathcal{Z}) =28$. We then apply the two-level preconditioner to the whole dataset.
The comparison of performances between the PCG run with $\mbd$ and $\md$ is shown in fig. \ref{fig:histos2D}, respectively with blue and orange bars. \gpedit{ Even in this case, we adopted the same tolerance,  $10^{-6}$, } and  we end the PCG iterations when this  tolerance  is not  achieved within 100 iterations. While most of the $\mbd$ runs do not converge, $\md$ runs converged within a median value of $ 44$ iterations.\\*

\begin{figure*}
\includegraphics[width=2\columnwidth]{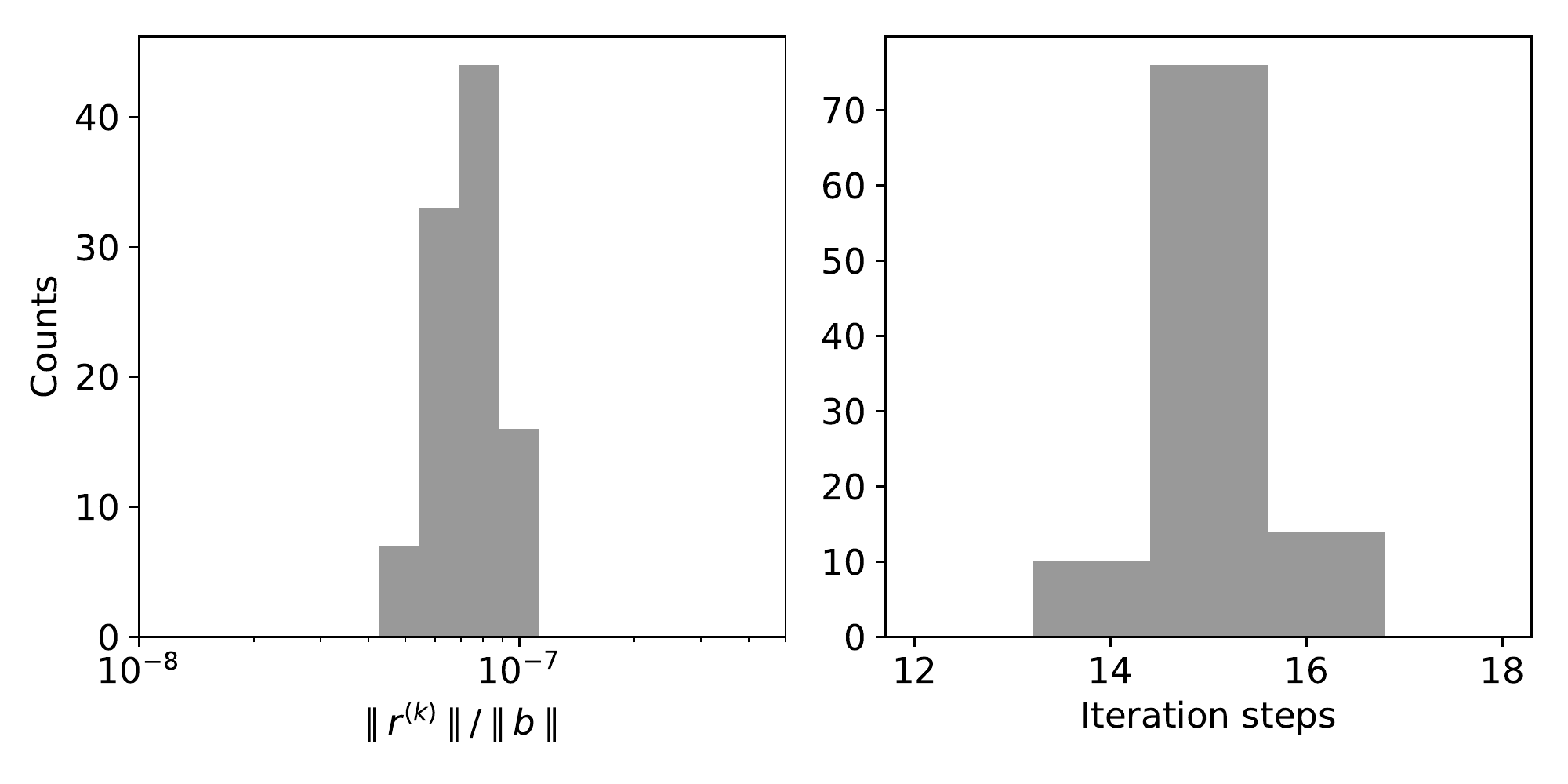}
\caption{PCG runs performed for 100 MC simulated data observing the VLP. Notice that all the 100 MC runs converged below a tolerance of $10^{-7}$ within $\sim  15$ iterations. }\label{fig:histo_mc}
\end{figure*}

The last case to be analyzed is the VLP, which targets $20\%$ of the sky with a sensitivity typical of future CMB observatories.  Similarly to the SP and LP case, we compare the performances of the two preconditioners applied to one year of signal-plus-noise observations -- a total of $300$ simulated 8-hour long CESs, grouped in 100 subsets of 3 CESs. We run the map-making solver on 100 processing elements distributed on 100 nodes of the \textsc{Cori}. 
Consistently with the previous LP case, we apply the $\md$ with the simplified approach with an \gpedit{ Arnoldi tolerance of $\epsilon_{Arn}=10^{-12}$.} In this case, the deflation subspace is spanned by 15 Ritz eigenvectors. As shown in Fig.~\ref{fig:histos2D}, also in this case while the $\mbd$ rarely \gpedit{ converges to the fixed tolerance ($10^{-6}$),  whereas $\md$  converges within few tens of iterations.}

Further details about the convergence statistics and total computational cost of SP, LP and VLP can be found in table \ref{tab:stats}. 

\subsection{Real space convergence}

We analyze the convergence performances of the two-level and block-diagonal preconditioners using the norm of the difference as defined in Eq.~\eqref{eq:errnorm}. Compared to the standard PCG residuals, this metric emphasizes more the eigenvectors of the system matrix with low eigenvalues. As mentioned earlier this analysis requires the knowledge of the exact solution of the system. For this reason, we perform  signal-only simulations for a  subgroup of all  the  observational patches discussed in sect. \ref{sec:divideConquerResults} and compare the performance of $\mbd$ and $\md$, computed with the  active approach.

The results, shown in Fig.~\ref{fig:resid_errs}, show that the two-level preconditioner is able to recover the solution to  some order of magnitude better precision with respect to  the one computed with the block-diagonal methodology. The fact that the latter saturates very quickly at a value way higher then the PCG residual emphasizes further that its convergence is hindered by the nearly degenerate modes, which are downweighted in the PGC residuals shown in the other plots, \eg\ Fig~\ref{fig:arntol_vs_dimZ}. Moreover, the fact that the  saturation levels differs case by case  in Fig.~\ref{fig:resid_errs}, could be  due to the presence of different degeneracies depending on the considered observational patch.

\subsection{Convergence at the power spectrum level}

We investigate a scale-dependence of the reconstructions by analyzing the signal-only study cases considered in the previous section and perform the bin-by-bin power spectra comparison of the residuals as shown in Fig~\ref{fig:spectrabins}. 

For the  SP case, the two-level preconditioner converges to the threshold of $10^{-7}$ within 40 iterations, whereas the case with $\mbd$  do not within 100 iterations, i.e., the maximum allowed in these runs. We consider the bins that are usually considered in the analysis of patches as small as the SP.  As one can notice from Fig.~\ref{fig:spectrabins} (top), the  solution computed with $\mbd $ encodes an extra-bias which is order of few percentages sub-dominant with respect to the variance of the signal itself, meaning that the quality of the map reconstructed with the $\mbd $ is acceptable as far as small angular scales are concerned. Moreover, this is somewhat expected since the larger angular scales are not constrained by the $\mbd$ and are the responsible of the long mode plateau we described in sect. \ref{sec:degen}.  Those scales are anyway unconstrained due to the small sizes of the patch.

LP allows us to probe larger scales, where the primordial gravitational wave B-mode signal is expected to peak. The solution computed with $\mbd$ (which does not converge within 100 iterations) shows a $\sim 10 \%$ bias at the largest angular scales (\ie\ in the first two bins, namely $\ell_b = 50-100, 150-200$ in Fig.~\ref{fig:spectrabins}(center panel)), whereas the bias is  not present into  the solution computed with $\md$. This result becomes even more remarkable given that at these scales the signal is likely to be dominated by foreground emission, therefore the same fractional bias in the power-spectrum can be comparable with the whole signal from primordial $B$-modes.  In terms of the norm of the standard residuals these results demonstrate that high precision convergence needs to be attained in order to ensure a sufficient precision of the recovered sky signal on all, and specifically on the largest accessible, angular scales.

We observe a similar behavior with the spectra computed for VLP, Fig.\ref{fig:spectrabins}(bottom). In particular, we focus onto  large scales since the size of the patch is big enough to probe the angular scales related to the \emph{reionization} peak of both $E$- and $B$-modes. We notice that the first two multipole bins $\ell_{b}=0-50$ and $\ell_{b}=50-100$ are reconstructed up to percentage level with the two-level preconditioner, whereas the power spectra computed with the block-diagonal one contains a bias which may fluctuate between tens and few percentages.  The degree and subdegree angular scales are similarly reconstructed as in the LP case.

\subsubsection{Monte Carlo simulations}

All modern CMB experiments produce or validate their statistical and systematic uncertainties using a large number of simulations. Typically, each of them solves for a map-making system that has the same system matrix $A$ but different RHS $b$ (i.e., the same scanning strategy and data processing but different synthetic time stream).
We consider an observation composed of 3 CESs covering the VLP. We produce 100 Monte Carlo (MC) with not only uncorrelated noise, but even a \cmb signal generated using different random seeds from the same CAMB power spectra. We take one of these simulations and build a two-level preconditioner from a fully converged Arnoldi run. We then apply the same preconditioner to all the simulations. As shown in Fig.~\ref{fig:histo_mc}, all the 100 MC runs converged to a residual tolerance $< 10^{-7}$ within $\sim 15$ iterations and with a staggering narrow dispersion. This result on one hand shows how powerful a two-level preconditioner can be when MC simulations are to be performed, on the other it means that the degeneracies preventing the convergence with the standard preconditioner are not due to the signal or the presence of noise, but mostly due to the scanning strategy and the filtering applied to the time stream.

\section{Summary and conclusions} \label{sec:conclusions}
In this work we described an implementation of a novel class of iterative solvers, the two-level preconditioners,  $\md$,~\citep{Grigori2012,2014A&A...572A..39S} in the context of the \cmb map-making procedure applied to data sets filtered at the time domain level. We have discussed the details of the construction of the new preconditioner and proposed a simplified, ''divide and conquer'', embarrassingly parallel implementation of the method, which can be adequate for an analysis of current and future, ground-based observations.
We have tested this new implementation of this novel methodology on three different simulated data sets in the cases when filtering operators typical of the ground experiments, have been applied. We have compared the performance of the method with that of the standard PCG solver based on the Jacobi preconditioner.

We have found that in all the studied cases, the two-level preconditioner, $\md$, have performed better both in terms of the attained precision and the number of required iterations, allowing typically reaching the residuals on order $10^{-7}$ within $20\div 40$ iterations. The standard approach yields residuals an order of magnitude or more higher within as many as $100$ iterations. We show that reaching such high precision of the reconstructed maps is required in order to constrain the  large angular scales of the B-mode polarization. Indeed, the new approach consistently produces maps typically within $20\div40$ iterations, which display negligible reconstruction bias of all and in particular the longest modes as represented in the maps. In the contrary, the maps derived with the standard solver with the maximal number of iterations set to $100$ show typically a $1-20 \%$ bias at all scales.\\ 
We thus conclude that producing highly accurate maps of the polarized CMB anisotropies from the filtered data of the ground-based experiments may call for more advanced iterative solvers than the standard PCG solver with the Jacobi preconditioner. The presented here, two-level preconditoner offers significantly better performance and could be a method of choice for such applications in the future. These advantages come however at the additional cost needed to construct the preconditioner. Therefore this method can be particularly useful in the cases of large MC simulations, where the additional cost is offset by the solver's superior performance.

\begin{acknowledgements}
We acknowledge use of {\sc camb}, {\sc healpix}, {\sc s4cmb}, and {\sc x2pure} software packages. We thank Julien Peloton for his help with simulations, and Josquin Errard and Maude Le Jeune for helpful comments and suggestions.
RS acknowledges support of the French National Research Agency (ANR) contract ANR-17-C23-0002-01 (project B3DCMB).\\
This research used resources of the National Energy Research Scientific Computing Center (NERSC), a DOE Office of Science User Facility supported by the Office of Science of the U.S. Department of Energy under Contract No. DE-AC02-05CH11231. GF acknowledges the support of the CNES postdoctoral program. CB, DP acknowledge support from the RADIOFOREGROUNDS project, funded by the European Commission’s H2020 Research Infrastructures under the Grant Agreement 687312, and the COSMOS Network from the
Italian Space Agency; CB acknowledges support by the INDARK INFN Initiative. \\
\end{acknowledgements}

\bibliographystyle{aa} 
\bibliography{pcg}

\begin{appendix}
\section{The Arnoldi Algorithm} \label{subsec:arnoldi}
\begin{figure}[!htbp]
\includegraphics[width=1.\columnwidth]{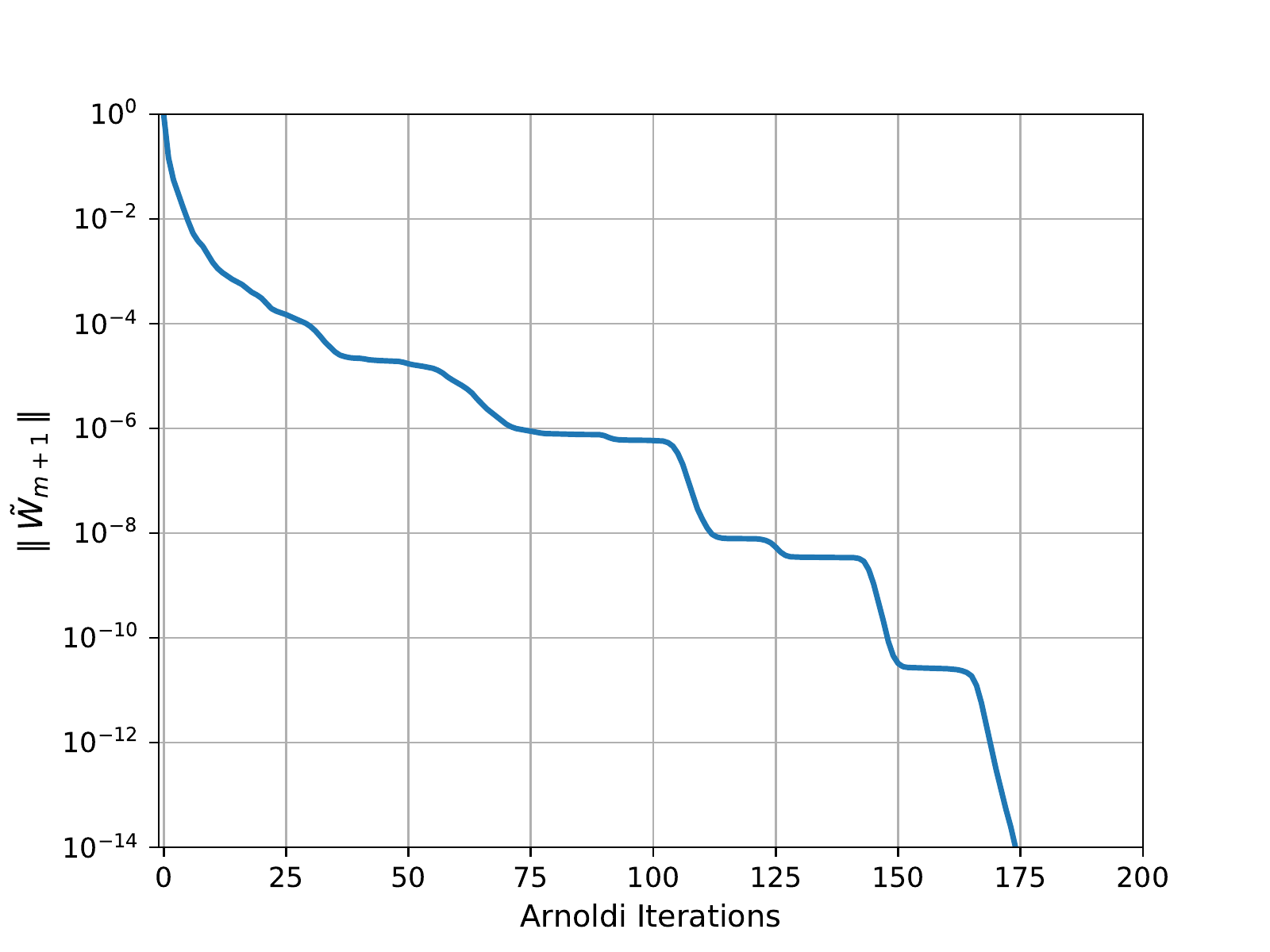}
\caption{The convergence residuals of the Arnoldi algorithm. Notice that after about $175 $ iterations the algorithm numerically converges.}\label{fig:arnoldires}
\end{figure}

\begin{figure*}[h!]
\includegraphics[width=2\columnwidth ,clip=true, trim=0 7.5cm 0 3cm ]{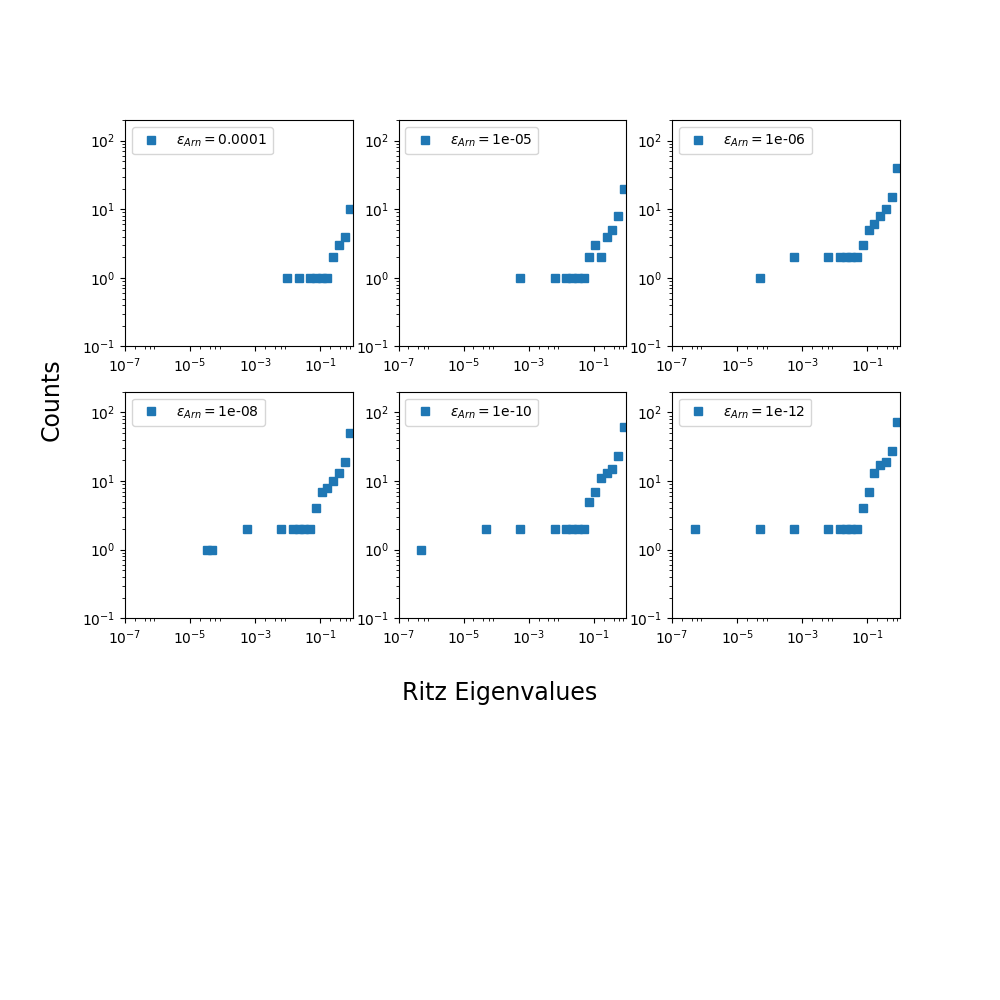}
\caption{Histograms of Ritz eigenvalues of the matrix $\mbd A$ computed for several choices of Arnoldi tolerance, $\epsilon_{Arn}$.}
\label{fig:binned_ritz}
\end{figure*}

The \emph{Krylov} subspace algorithms are based on the construction of a sequence of vectors naturally produced by the power method,  a class of  those, namely the  Minimal Residuals (MINRES) and the Generalized Minimal Residual (GMRES) methods \citep{Golub:1996:MC:248979}, rely on the Arnoldi algorithm. 
Our goal is to find an approximation to the eigenvalues of a matrix $B$ of  a generic linear system with RHS $b$:
\begin{equation}
Bx=b.
\label{eq:newlinsys}
\end{equation}
The Arnoldi algorithm is an algorithm aimed at solving linear systems by projecting the system matrix onto a convenient Krylov subspace generated by the first $m$ vectors
\begin{equation}
\mathcal{K}_m (B, b)= span \lbrace b, Bb,B^2 b, \dots, B^{m-1} b \rbrace.
\label{eq:krylov}
\end{equation}
The  major steps of the algorithm are summarized in Algorithm~\ref{alg:arnoldi}.
\begin{algorithm}[H]
\caption{Basic steps of the Arnoldi algorithm}\label{alg:arnoldi}
  \begin{algorithmic}[1]
    \Require : $r_0,\, w_1= r_0/\parallel r_0 \parallel$
    \For{j = 1$\rightarrow $ m}
    	\State $v_j=B w_j$
    \For{i = 1$\rightarrow $ j}
      \State $h_{i,j}= (v_j , w_i)$
    \EndFor
	\State    $v_j = v_j - \sum_{i=1} ^j h_{i,j} w_i$ 
	\State $h_{j+1,j} = \parallel  v_j\parallel$
	\If{$h_{j+1,j}\leq \mathrm{tol}_{Arn}$  }
		\State break
	\Else 
		\State $w_{j+1} = v_j / h_{j+1,j}$
	\EndIf
	\EndFor
  \end{algorithmic}
\end{algorithm}
\noindent
Hence, the output of the Arnoldi algorithm is an orthonormal basis $W^{(m)}= (w_1| w_2|\dots| w_m)$ (called the \emph{Arnoldi vectors }), together with a set of scalars $h_{i,j}$ (with $i,j=1,\dots,m$ and $i\leq j+1$) plus an extra-coefficient $h_{m+1,m}$. The former set of coefficients are the elements of an upper Hessenberg matrix $H_m $ with non-negative subdiagonal elements and is commonly referred as a \emph{m-step Arnoldi Factorization} of $B$. If $B$ is Hermitian then $H_m$ is symmetric, real  and  tridiagonal and the vectors (columns of $W^{(m)}$) of the Arnoldi basis  are called \emph{Lanczos vectors}. $B$ and $H_m$ are intimately related via:
\begin{equation}
BW^{(m)} = W^{(m)} H _m + h_{m+1,m} w_{m+1} e^\dag  _m,
\label{eq:factoriz}
\end{equation}
where $e_m$ is a $1\times m $ unit vector with 1 on the $m$-th component.
In other words, $H_m$ is the projection of $B$ onto the subspace generated by the Arnoldi basis $W^{(m)}$  within an error  given by $\tilde{W}_m= h_{m+1,m} w_{m+1} e^\dag  _m$.  The  iteration loop ends when this error term gets smaller than a  certain threshold $\epsilon_{Arn}$. \\*
Using Eq.~\eqref{eq:factoriz}, we can connect the eigenpairs of $B$ to the ones of $H_m$.
Let us consider an eigenpair of $H_m, \,(\lambda_i,y_i)$
\begin{displaymath}
H_m y_i = \lambda_i y_i.
\end{displaymath}
The vector $v_i= W^{(m)} y_i$ then satisfies
\begin{equation}	
\parallel B v_i - \lambda_i v_i \parallel= \parallel ( B W^{(m)} - W^{(m)} H_m ) v_i \parallel=  \parallel \tilde{W}_m v_i \parallel.
\label{eq:ritz}
\end{equation}
The eigenpairs of $H_m$ are therefore approximations of the eigenpairs of $B$ within an  error given by $\tilde{W}_{m+1}$. They are the so called Ritz eigenpairs  and they are very easy to compute since the size of $H_m$ is  $\lesssim \mathcal{O}(100)$. For the \cmb dataset considered in this work this is indeed the order of Arnoldi iterations required to reach a tolerance $\epsilon_{Arn}\sim 10^{-6}$ as it can be seen in Fig.~\ref{fig:arnoldires}. A typical distribution of the amplitude of the Ritz eigenvalues for different values of $\epsilon_{Arn}$ is shown in in Fig.~\ref{fig:binned_ritz}.
\begin{figure}[h]
\includegraphics[width=1\columnwidth]{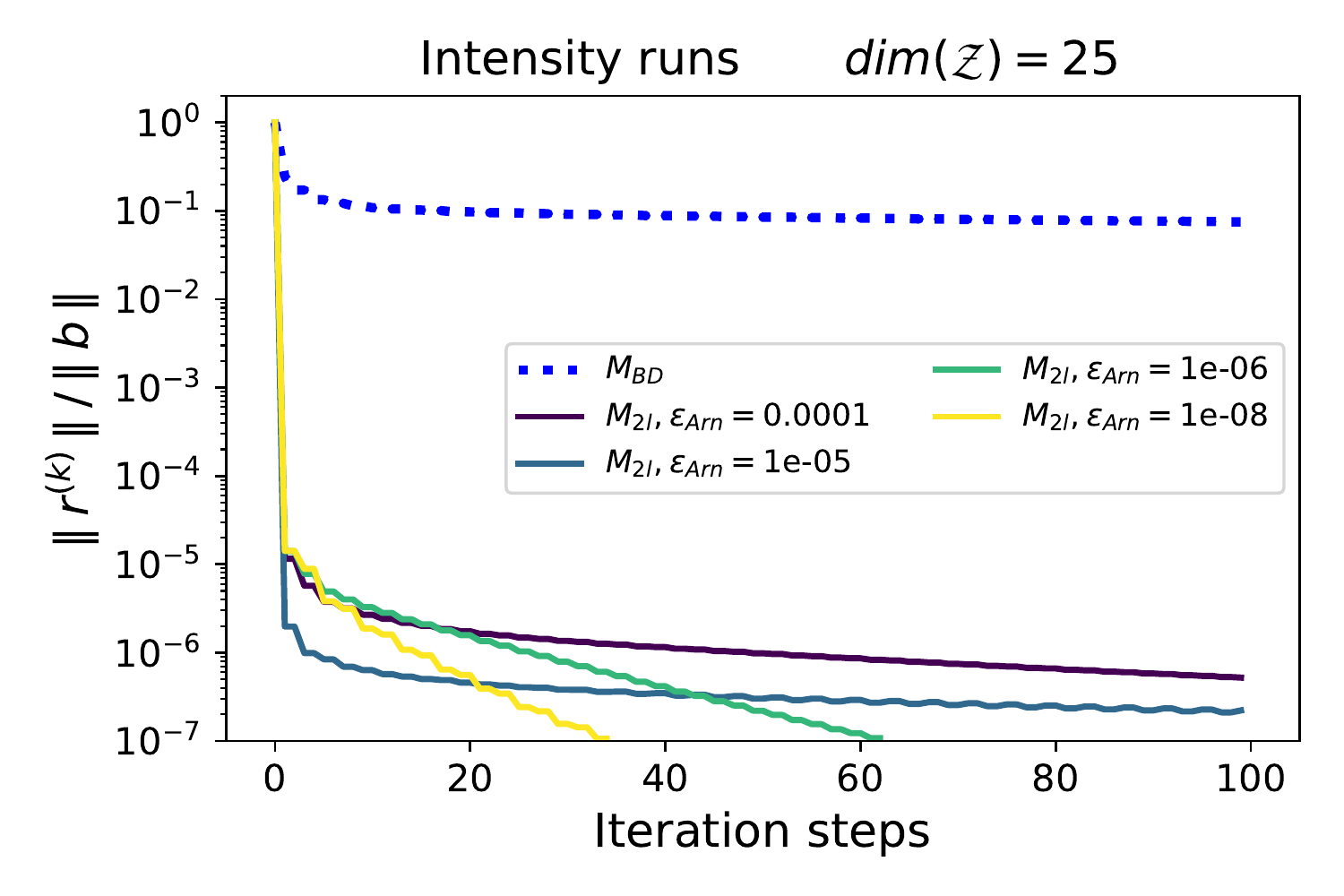}
\caption{PCG residuals for SP intensity maps different choices of $\epsilon_{Arn}$  and at a given $\mathrm{dim} \mathcal{Z}= 25$.}\label{fig:T_pcg}
\end{figure}
\section{Solving for intensity maps}\label{app:Truns}

Similarly to what we have done in Sect.~\ref{sec:constr2lvl}, we further tested the two-level preconditioner on SP intensity-only maps. As it is shown in Fig.~\ref{fig:psd}, in this case the time stream has to be filtered with  a higher  polynomial basis due to a larger  $f_{knee}$. We therefore filter the time streams up to the third order Legendre polynomials. 
Fig.~\ref{fig:T_pcg} shows the PCG residuals for different choices of $\epsilon_{Arn}$ and 	one can easily notice the similarity to the right panel of Fig.~\ref{fig:arntol_vs_dimZ}. This further indicates that our results are stable even when a more aggressive filter is applied to the data. Moreover, by looking at the blue-dashed line in  Fig.~\ref{fig:T_pcg}, the $\mbd$  residuals saturate at a higher threshold with respect to the polarization case (Fig.\ref{fig:arntol_vs_dimZ}), remarking  the presence  of different degeneracies present when intensity maps are involved. However, the two-level preconditioner does not suffer of this effect and once the Ritz eigenvector basis is very well approximated, by running the Arnoldi algorithm to tolerances below $10^{-6}$,   it converges to $10^{-7}$ within $\sim 40$ iterations. 

\end{appendix}

\end{document}